\documentclass[fleqn,usenatbib]{mnras}

\usepackage[T1]{fontenc}

\DeclareRobustCommand{\VAN}[3]{#2}
\let\VANthebibliography\thebibliography
\def\thebibliography{\DeclareRobustCommand{\VAN}[3]{##3}\VANthebibliography}

\usepackage{graphicx}	
\usepackage{amsmath}	
\usepackage{amssymb}	

\usepackage{newtxtext,newtxmath}
\usepackage{bm}
\usepackage{epstopdf}
\usepackage{verbatim}
\usepackage{natbib}
\usepackage{hyperref}
\usepackage{float}
\usepackage{color}
\usepackage{footnote}
\usepackage{threeparttable}%

\title[Proper Motions of 3C 78 with the VLA]{\textit{Proper Motions in the sub-kiloparsec Jet of 3C 78: Novel Constraints on the Physical Nature of Relativistic Jets}}

\author[A. Roychowdhury et al.]{
Agniva Roychowdhury,$^{1}$\thanks{E-mail: agniva.physics@gmail.com}
Eileen T. Meyer,$^{1}$
Markos Georganopoulos,$^{1,2}$
and Kassidy Kollmann$^{3}$
\\
$^{1}$Department of Physics, University of Maryland Baltimore County, 1000, Hilltop Circle, Baltimore MD 21250, USA\\
$^{2}$NASA Goddard Space Flight Center, Code 663, Greenbelt, MD 20771, USA\\
$^{3}$Department of Physics, Princeton University, Princeton, NJ 08544, USA\\
}

\date{Accepted XXX. Received YYY; in original form ZZZ}

\pubyear{2023}

\begin{document}
\label{firstpage}
\pagerange{\pageref{firstpage}--\pageref{lastpage}}
\maketitle

\begin{abstract}

Jets from active galactic nuclei are thought to play a role in the evolution of their host and local environments, but a detailed prescription is limited by the understanding of the jets themselves. Proper motion studies of compact bright components in radio jets can be used to produce model-independent constraints on their Lorentz factor, necessary to understand the quantity of energy deposited in the inter-galactic medium. We present our initial work on the jet of radio-galaxy 3C~78, as part of CAgNVAS (Catalogue of proper motions in Active galactic Nuclei using Very Large Array Studies), with a goal of constraining nature of jet plasma on larger ($>100$ parsec) scales. In 3C~78 we find three prominent knots (A, B and C), where knot B undergoes subluminal longitudinal motion ($\sim0.6c$ at $\sim$ 200 pc), while knot C undergoes extreme (apparent) backward motion and eventual forward motion ($\sim-2.6c$, $0.5c$, at $\sim$ 300 pc). Assuming knots are shocks, we infer the bulk speeds from the pattern motion of Knots B and C. We model the spectral energy distribution (SED) of the large-scale jet and observe that a physically motivated two-zone model can explain most of the observed emission. We also find that the jet profile remains approximately conical from parsec to kiloparsec scales. Using the parsec-scale speed from VLBI studies ($\sim0.1c$) and the derived bulk speeds, we find that the jet undergoes bulk acceleration between the parsec and the kiloparsec scales providing the first direct evidence of jet acceleration in a conical and matter-dominated jet.

\end{abstract}




\section{\rm introduction}

Bipolar kiloparsec-length jets of relativistic plasma are launched from about $\sim5\%$ of active galactic nuclei (AGN) \citep{bland79,begelman84,urry95,blandford2019}. While high-resolution multi-wavelength studies have provided numerous looks at these objects for the last few decades \citep[e.g.,][]{fan74,sch00,allen02,harris06,temb09,massaro15,blandford2019,hardcast20}, the fundamentals of how these jets are formed and accelerated \citep[e.g.,][]{blandford77,bland82,komis07,tch11}, how they are composed \citep[e.g.,][]{georg05,kat08,mehta09}, how they propagate \citep[e.g.,][]{blandford90,falle91,gaibler09,mckinney09}, and how they affect their large-scale environment through feedback \citep[e.g.,][]{fabian12,king15} are still debated. Present studies are generally plagued by large uncertainties and bias in the vital questions, such as measures of the energy content of jet, effects of the jet viewing angle and the physical nature of the plasma when we only observe the emitted radiation.

Studies of proper motions of "knots", or bright compact inhomogeneities at different positions in these jets can, in principle, be used to produce model-independent constraints on the Lorentz factor ($\Gamma$) of the bulk motion at different spatial scales in the jet, when carefully combined with the jet-counterjet flux ratio. This inferred velocity profile of the jet, in addition to constraining theories of jet acceleration-deceleration, is vital in understanding the extent of kinetic feedback at the cluster environment or the evolution of the jet composition as it propagates. It is possible to partly constrain the latter from bulk Comptonization models \citep[e.g.][]{georg05,mehta09} and models of the Compton drag effect \citep{sikora96,rivas14} using $\Gamma(z)$, where $z$ is the distance along the jet from the black hole.

Proper motion studies of these jets exist in plenty at the parsec and sub-parsec scales, observed as a part of many VLBI monitoring programs by different groups \citep[e.g.,][]{reid89,junor95,zensus95,tingay98,jorstad01,homan01,middel04,muller14_tanami,lister16,piner18,walker18}. These studies have produced strong statistical constraints on the velocity, structure, magnetic field and dynamical evolution of a sample of $\sim$ 500 parsec-scale jets. In contrast, studies of these jets on the larger $>100$ parsec scales are very rare, majorly owing to the requirement of large time baselines ($\gtrsim$ few decades) to achieve appreciable accuracy. The jet kinematics of M87 are currently the most well studied, using proper motion results from the Very Long Baseline Interferometry (VLBI), Hubble Space Telescope (HST), Chandra and the Very Large Array (VLA), where the knots, initially slow ($\lesssim 0.3c$) at the parsec scales, were found to reach speeds of $\sim6c$ at the site of HST-I, before eventually decelerating \citep{reid89,biret95_hst,biret99,cheung07,kovalev07,ly2007,meyer13,asada14,snios19}. Furthermore, knots in the large-scale jets of 3C 264 \citep{meyer15}, 3C 273 \citep{meyer16_3c273}, M84 \citep{meyer18_m84}, Cen A \citep{snios_cenA} and 3C 120 \citep{walk97}, observed by the HST, VLA or Chandra, were found to exhibit very slow sub-luminal to superluminal motions. In addition to M87, clear evidence of acceleration of bulk flow at the hundred-parsec scales has been obtained for both 3C 264 and Cen A. 

These studies have ushered in a new era of understanding the nature of large-scale jets. The "standard model" of magnetic acceleration was proposed to be at work in the jet of M87 \citep[e.g.,][]{kov20}, where magnetic energy is continually expended to propel the jet \citep[e.g.,][]{komis07}, marked by a transition in parabolic to conical shapes between the Poynting-dominated and the matter-dominated regimes. For M84, 3C 273 and Cen A, constraints on the bulk velocity at the $>100$ parsec scales have enabled a better understanding of their spectral energy distributions, and in turn the underlying emission processes. For example, the widely used IC/CMB \citep[e.g.,][]{cel01,tav00} model used to explain bright X-ray emission from kiloparsec-scale jets needs highly relativistic velocities, which could be disproven in the kpc-scale jet of 3C 273 through approximate model-independent measurements of speed and careful spectral fitting \citep{meyer16_3c273}.

The motivation for studies like these have hugely increased since the advent of the VLA since the time baselines are $>40$ years, increasing accuracy in measuring proper motions. The VLA archives are rich, containing thousands of observations of radio-loud AGN, at various bands and resolutions. Using the richness of the VLA archive and continually proposed observations, our goal is to create a large Catalogue of proper motions in Active galactic Nuclei using Very Large Array Studies (CAgNVAS), with a view to investigating physical properties of large-scale jets, both statistically and on a case-by-case basis. In \cite{royc_evn23, royc23_arxiv_ngdif}, we have discussed the detailed radio-interferometric techniques required to accurately constrain velocities of components in large-scale jets. In this paper, we have used those techniques on radio-loud AGN, 3C 78, and have constrained various physical properties of its jet.

3C 78 (J0308+0406) is a nearby ($z=0.029$) broad-lined radio galaxy with an $\sim$ arcsecond long (1"/0.58 kpc, at $H_0\sim70$ km s$^{-1}$ Mpc$^{-1}$) jet, whose multi-wavelength properties have been moderately studied in the radio to the hard X-rays \citep[e.g.,][]{unger84,colina90,liu92,sparks95,zirbel98,martel99,balma12,massaro15} at different resolutions. It has been specifically monitored as a MOJAVE target, where the maximum observed $\beta_{\rm app}\sim0.1$ at the $\sim$ parsec scale \citep{lister18}. In addition to proximity, the target has a very well-defined large-scale jet in both the radio \citep[e.g.,][]{unger84,saik86} and optical \citep{chia02}, thereby making it an ideal test-bed to compare proper motions at different wavelengths and thereby investigate the nature of knots and their connection to the bulk motion. We have organized the paper as follows. We discuss the data reduction in Section \ref{sec:data}. In Section \ref{sec:pm}, we discuss the data analysis techniques to determine proper motions as a function of distance along the jet. In Section \ref{sec:disc}, we discuss the implications of our results keeping an eye to the jet spectral energy distribution. In Section \ref{sec:conc}, we conclude with a summary of our results.

\section{Data Reduction}
\label{sec:data}
3C 78 has been observed by the Chandra X-ray Observatory and we obtained X-ray fluxes of the core from and the jet from \cite{massaro15} and \cite{fukazawa15}. 3C 78 also has a bright optical jet \citep{sparks95} observed by the the Hubble Space Telescope (HST) for multiple epochs over a baseline of $\sim$ 25 years but we aim to discuss this in a future paper. The optical core (dominated by jet emission) and jet fluxes were obtained from \cite{chia02} and \cite{sparks95} respectively, while the Fermi LAT $\gamma$-ray fluxes from \cite{fukazawa15}. They have been listed in Table \ref{table:other_fluxes}. Here we describe the radio data reduction procedure of 3C 78.

\subsection{Very Large Array (VLA)}

\begin{figure}
    \centering
    \includegraphics[width=0.85\linewidth]{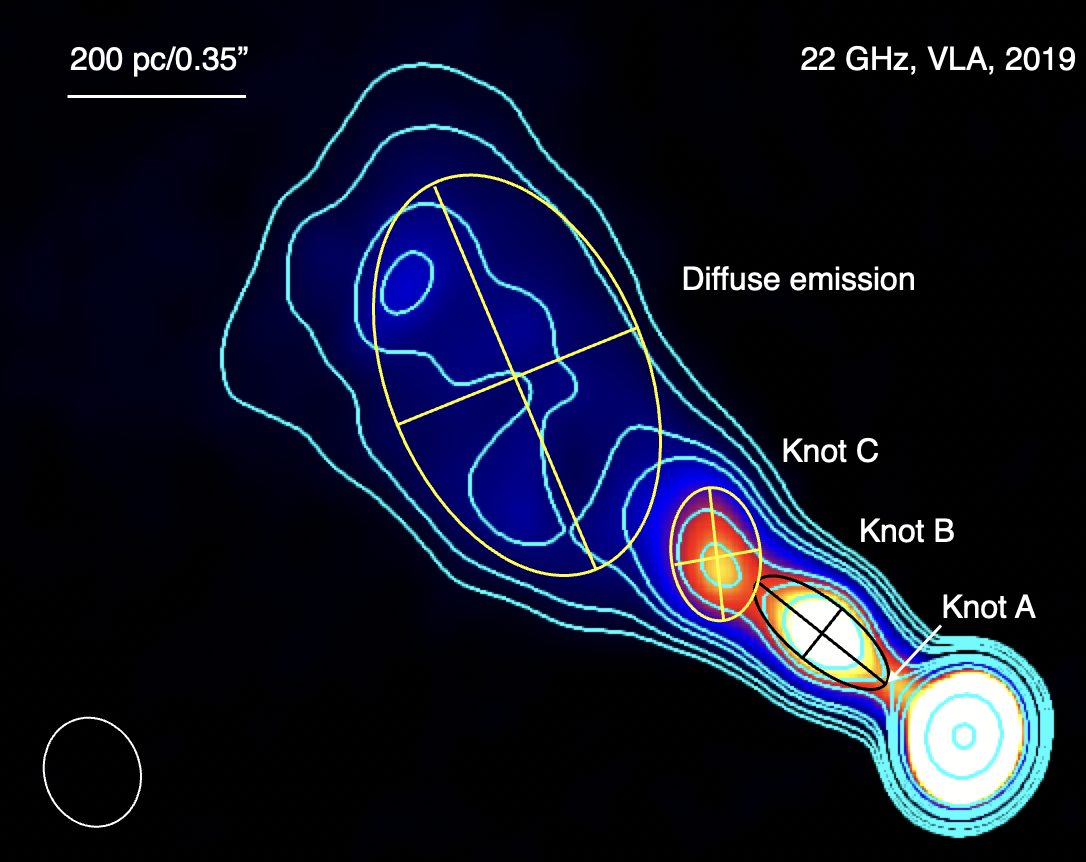}
    \caption{22 GHz 2019 VLA image of 3C 78. Three compact knots are visible in addition to the core, which have been labelled accordingly. Knots B and C are more prominent and have been described by a Gaussian ellipse fit (from \texttt{DIFMAP}). The diffuse emission further away has also been modelled by a Gaussian, shown in yellow. The contour levels are as follows (in mJy/beam): 0.2, 0.35, 0.7, 1, 2.5, 3.5, 5, 10, 100, 300, 500.}
    \label{fig:vla19}
\end{figure}

We reduced six VLA observations of 3C 78 using the Common Astronomy Software Applications \citep[CASA;][]{casa}, from 1985 to 2019, for studying both proper motions and the spectral energy distribution. The observations for proper motions were chosen on the basis of high resolution (of the order $\sim0.1"-0.2"$), suitable for tracking compact jet components which are otherwise blended in lower resolution observations. We have listed a summary of these observations with the relevant image properties including peak/core and jet flux densities, final image RMS in Jy, synthesized beam size in arcsec and largest angular scale in Table \ref{table:vla}. The proper motion datasets were AM141, AF376, AP439 and 19A-168, while the remaining were only for measurements of total jet/unresolved core flux. We applied standard manual calibration to the classic VLA (1985, 1988, 2000, 2003) datasets. For the 2019 JVLA observation, we used CASA pipeline version 5.4.0. Since 3C 78 is core-dominated and very bright in all radio imaging, multiple rounds of phase and amplitude self-calibration were applied to improve the final imaging using \texttt{clean}, where we used Briggs weighting with robustness=0.5 \citep{briggs99}, to maintain equal balance between sensitivity and resolution.

For the spectral energy distribution, we used the peak flux as the sub-kpc core flux, while we produced core-subtracted images to more accurately measure the total jet flux. To this end, we applied the \texttt{clean} deconvolution task in CASA to create a point source model for only the core and then subtracted it from the total visibility data using CASA task \texttt{uvsub}. We followed this by a final \texttt{clean} of the hence core-subtracted visibility to produce a final image without the core. The total jet flux density for each of the VLA datasets was then equal to the flux density of a large region containing all of the extended jet emission.

Figure \ref{fig:vla19} shows 3C 78 at 22 GHz. It extends through $1"/580$ pc and majorly has 3 very bright compact components labelled A, B and C, in addition to an extended outer jet further away from Knot C. In this paper, we discuss the structural and positional evolution of Knots B and C with time, in order to place an indirect constraint on the bulk flow characteristics of the jet.

\begin{figure*}
    \centering
    \includegraphics[scale=0.25]{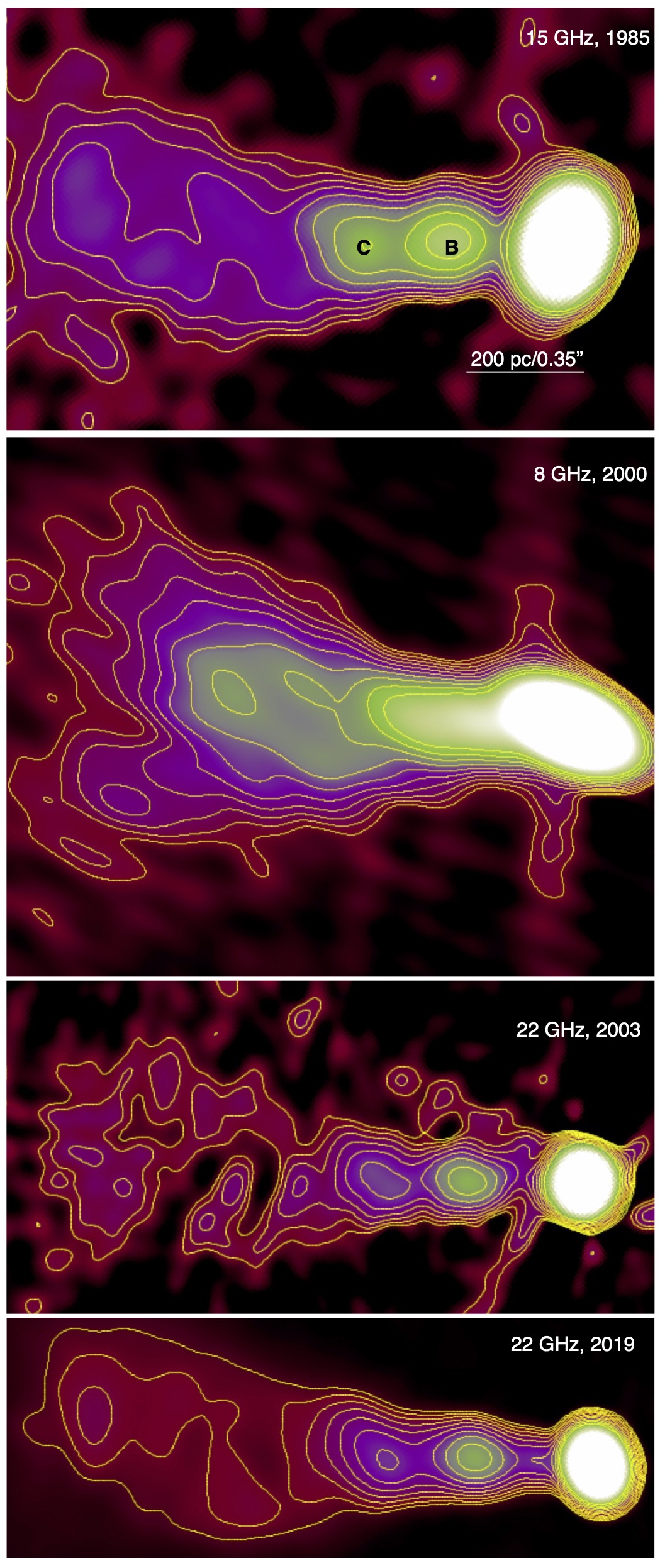}
    \caption{Multi-epoch images of 3C 78 showing Knots B and C. The images have been rotated by the position angle of the jet direction for better image comparison and compactness. Different panels have different resolutions and contour levels and are described as follows, in Jy/beam. The values were chosen by trial and error to emphasize the brightest pixel in each knot and the faintest pixel at the jet boundary to the maximum extent. Top panel: 0.001 to 0.016 in factors of $\sqrt{2}$. Second panel: 0.0007 to 0.0160 in factors of $\sqrt{2}$. Third panel: 0.0007 to 0.0160 in factors of $\sqrt{2}$. Bottom panel: 0.0005 to 0.0160 in factors of $\sqrt{2}$.}
    \label{fig:vla_all}
\end{figure*}

\begin{table*}
\centering
\caption{\label{table:other_fluxes} Optical, X-ray and $\gamma$-ray upper limits/fluxes.}
\begin{tabular}{ccccc}\hline
Instrument & Epoch & Frequency ($\nu$) & Frequency times Flux ($\nu F_\nu$) & Reference \\
         & (MM/YY) & (Hz) & (erg s$^{-1}$ cm$^{-2}$ Hz) & \\
\hline
HST & 03/00 & $10^{13.67}$ & Core, $10^{-11.67}$ & \cite{chia02} \\
 & 03/00 & $10^{14.12}$ & Core, $10^{-11.65}$ & \\
& 08/94 & $10^{14.64}$ & Jet, $10^{-12.84}$ & \cite{sparks95} \\
Suzaku & 09/09 & $10^{16.19}$ & Core, $10^{-11.91}$ & \cite{fukazawa15} \\
 & & $10^{16.73}$ & Core, $10^{-12.11}$ & \\
Chandra & 06/04 & $10^{17.26}$ & Core, $10^{-12.29}$ & \cite{massaro15} \\
& & $10^{17.26}$ & Jet, $10^{-13.81}$ &  \\
& & $10^{17.56}$ & Core, $10^{-12.18}$ &  \\
& & $10^{17.56}$ & Jet, $10^{-13.65}$ &  \\
& & $10^{18.04}$ & Core, $10^{-12.00}$ &  \\
& & $10^{18.04}$ & Jet, $10^{-13.41}$ &  \\
Fermi-LAT & 2008-2013 & $10^{22.67}$ & $10^{-12.06}$ & \cite{fukazawa15} \\
& & $10^{23.17}$ & $10^{-11.84}$ & \\
& & $10^{23.64}$ & $10^{-12.11}$ & \\
& & $10^{24.18}$ & $10^{-12.01}$ & \\
& & $10^{24.69}$ & $10^{-12.07}$ & \\
\hline
\end{tabular}
\end{table*}

\begin{table*}

\caption{Radio observations. RMS : root mean squared sensitivity, P.A. : position angle and LAS : largest angular (recoverable) scale. Length refers to the total on-source integration time in hours.}
\begin{tabular}{llclllccccccl} \hline\hline 
Obs. & Array & Freq.  & Project  & Date       & Length & Bandwidth & Beam  & P.A. &  LAS         &  RMS               & $F_\mathrm{core}$ & $F_\mathrm{jet}$ \\
            &    Config  &    (GHz)       &               & {\footnotesize YYYY-MM-DD} & (hr) & (MHz) &           (arcsec)  & (deg) & (arcsec)             & (Jy)              &  (Jy)    & (Jy)         \\
\hline
VLA         &  A  &   1.5        &     AC243 & 1988-12-31 & 0.08 & 50.0 & $1.53\times1.31$ & -6.2 & 36.0 & $2.9\times10^{-3}$      &   0.794 & 0.301 \\
VLA         &  A  &   4.8        & AR116 & 1985-01-08 & 0.05 & 50.0 & $0.55\times0.40$ & 34.7 & 29.0 & $2.5\times10^{-2}$      &   0.832 & 0.151 \\ 
VLA         &  A   &   8.4        & AF376         & 2000-12-10 & 0.20 & 50.0 & $0.28\times0.15$ & 31.0 & 17.0 & $1.2\times10^{-4}$ & 0.616 & 0.182 \\
VLA         &  A   &   15.0         & AM141         & 1985-02-19 & 0.18 & 50.0 & $0.16\times0.12$ & -49.7 & 12.0 & $3.3\times10^{-4}$ & 0.689 & 0.145 \\
VLA         &  A   &   22.0         &    AP439 & 2003-09-06 & 5.00 & 50.0  &$0.10\times0.09$ & 25.1 & 7.9 &   $9.1\times10^{-5}$      &   0.504 & 0.087 \\
VLA         &  A  &   22.0        &    19A-168  & 2019-08-08 & 0.30 & 128.0 & $0.11\times0.08$ & -14.4 & 7.9 & $3.4\times10^{-5}$      &   0.541 & 0.094 \\
ALMA        &    &   88.9        &    2018.1.00585  & 2018-11-26 & 0.50 & 1875.0 & $0.85\times0.74$ & 79.7 & 11.5 &   $6.6\times10^{-3}$      &   0.351 & 0.030 \\
\hline
\end{tabular} 
\label{table:vla}
\end{table*}

\subsection{Atacama Large Millimeter/submillimeter Array (ALMA)}

At resolutions comparable to the VLA data, 3C 78 has only been observed once by ALMA and the corresponding observational details have been tabulated in Table \ref{table:vla}. We used the appropriate CASA pipeline version (5.4.0) to calibrate the data and prepare the measurement set (MS) for imaging with \texttt{clean}. We used several rounds of (non-cumulative) phase-only self-calibration and a final amplitude and phase self-calibration to improve the sensitivity and dynamic range of the final image. To calculate the total jet flux for the spectrum, we followed a procedure similar to that described for the VLA images. Since the 89 GHz resolution is $\sim3-4$ times lower than the VLA X band observation (see Table \ref{table:vla}), we were careful to leave out regions beyond the X-band jet length of 3C 78. Particularly, the 89 GHz ALMA jet is $\gtrsim1"$ longer than the X band jet and contains an excess of around $\sim0.02$ Jy which is not detected in the X band.

\section{Analyzing VLA Data for Proper Motions}
\label{sec:pm}
\subsection{Model-fitting VLA Data}

\begin{table*}
\caption{\label{table:bestpar} Best-fit parameters and components for all epochs. All the components are Gaussians, except for the core model of 2000, where we deemed a point source was a better fit. X and Y positions denote the distances in the Eastern and Northern directions from the phase tracking centre. FWHM refers to the Full Width at Half Maximum and P.A. is the position angle, which measures an effective tilt of the major axis from the vertical at the phase centre.}
\begin{tabular}{cccccccccc} \hline \hline
Component & Epoch & Freq. & Resolution & Flux Density & East (X) & North (Y) & Major FWHM & Minor FWHM & P.A. \\
         &  & (GHz)  & (mas) & (Jy)  & (mas) & (mas) & (mas) & (mas) & (deg) \\ 
\hline
Core & 1985 & 15 & 130 & 0.666 &  -0.3 & 3.0 &  13.5 & 9.9  & -74.97 \\
& 2000 & 8 & 200 & 0.615 & -0.8 & -0.8  & & & \\
& 2003 & 22 & 90 & 0.514 & -0.1 & -0.1  & 6.9 & 3.7 & -90.00 \\
& 2019 & 22 & 90 & 0.561 &  4.9  &  3.8  &  13.3 & 1.2 & 2.3 \\
Knot A & 2003 & 22 & 90 & 0.007 & 124.3 & 97.6 & 263.1 &  0.0 & 49.49 \\
& 2019 & 22 & 90 & 0.004 &  124.6 &  85.2 &  78.9 & 0.0 & 42.2 \\
Knot B & 1985 & 15 & 130 & 0.041 &  260.9 &  197.6 &  184.4 & 58.6 & 52.04 \\
& 2000 & 8 & 200 & 0.038 &  262.0 &  195.4  &  126.4 & 61.7 & 47.06 \\  
& 2003 & 22 & 90 & 0.025 & 267.9 & 199.2 & 125.8 & 72.2 & 41.12  \\
& 2019 & 22 & 90 & 0.027 &  272.7 & 202.7 &  142.0 & 79.2 & 53.5 \\
Knot C & 1985 & 15 & 130 & 0.027 &  491.5 &  351.3 &  210.9 & 158.2 & 28.84 \\
& 2000 & 8 & 200 & 0.041 &  469.3 &  341.4 &  266.5 & 141.7 & 38.42 \\
& 2003 & 22 & 90 & 0.021 & 469.2 & 339.7 & 229.7 & 128.1 & 34.58 \\
& 2019 & 22 & 90 & 0.020 &  479.4 &  346.1 &  217.9 & 133.7 & 35.1 \\
Diffuse emission & 1985 & 15 & 130 & 0.072 &  920.6 &  731.7 &  796.8 & 487.1 & 33.28 \\
& 2000 & 8 & 200 & 0.102 &  953.5 &  748.5 &  887.3 & 582.1 & 32.60 \\
& 2003 & 22 & 90 & 0.049 & 915.6 & 727.8 & 854.0 & 503.7 & 35.30 \\
& 2019 & 22 & 90 & 0.052 &  920.6 &  764.8 &  922.0 & 502.6 & 31.2 \\
\hline
\end{tabular}
\end{table*}

\begin{table}
\centering
\caption{\label{table:gainerr} Estimated amplitude and phase errors.}
\begin{tabular}{ccccc}\hline
Epoch & Band/Config & Frequency & Amplitude Error & Phase Error \\
         & & (GHz) & (\%) & (radians)  \\ 
\hline
1985 & Ku/A & 15.0 & 8\% & 0.08 \\
2000 & X/A & 8.0 & 5\% & 0.05 \\
2003 & K/A & 22.0 & 20\% & 0.20 \\
2019 & K/A & 22.0 & 10\% & 0.10 \\
\hline
\end{tabular}
\end{table}

\begin{table*}
\caption{\label{table:pospar} Final determined X and Y positions, relative to the core, for Knots B and C. The longitudinal and transverse positions have been determined with respect to the 1985 jet axis.}
\begin{tabular}{ccccccccccc} \hline \hline
Knot & Epoch & East ($X-X_0$) & North ($Y-Y_0$) & Longitudinal & Transverse \\
         & & (mas) & (mas) & (mas) & (mas)  \\ 
\hline
A & 1985 & 261.2$\pm2.5\sqrt{2.0}$ & 194.6$\pm2.5\sqrt{2.0}$ & 325.7$\pm$2.8 & 0.0$\pm$0.0 \\
& 2000 & 262.8$\pm2.0\sqrt{2.0}$ & 196.2$\pm2.0\sqrt{2.0}$ & 328.0$\pm$2.3 & 0.3$\pm$4.6 \\
& 2003 & 265.0$\pm2.5\sqrt{2.0}$ & 198.3$\pm2.5\sqrt{2.0}$ & 331.0$\pm$2.8 & 0.7$\pm$5.1 \\
& 2019 & 267.7$\pm2.0\sqrt{2.0}$ & 199.9$\pm2.0\sqrt{2.0}$ & 334.1$\pm$2.3 & 0.4$\pm$4.6 \\
B & 1985 & 491.8$\pm4.0\sqrt{2.0}$ & 348.3$\pm4.0\sqrt{2.0}$ & 602.6$\pm$4.6 & 0.0$\pm$0.0 \\
& 2000 & 470.1$\pm3.5\sqrt{2.0}$ & 342.2$\pm3.5\sqrt{2.0}$ & 581.4$\pm$4.0 & 7.6$\pm$7.3 \\
& 2003 & 469.3$\pm5.0\sqrt{2.0}$ & 339.8$\pm5.0\sqrt{2.0}$ & 579.4$\pm$5.7 & 6.1$\pm$8.8 \\
& 2019 & 474.5$\pm3.0\sqrt{2.0}$ & 342.3$\pm3.0\sqrt{2.0}$ & 585.1$\pm$3.5 & 5.1$\pm$6.9 \\
\hline
\end{tabular}
\end{table*}

Figure \ref{fig:vla_all} shows the multi-epoch and multi-frequency VLA images of 3C 78, with the 22 GHz 2019 observation displayed separately in Figure \ref{fig:vla19}. Through close inspection of the images of the four epochs of VLA data, we found that the source structure is best described by a point source core (narrow Gaussian) with three or four other Gaussians (knots), depending on the resolution. However, the motions of Knots B and C are unclear across different \textit{images} of epochs due to the very slow speeds and different resolutions involved.

Analysis of arc-second scale VLA data to determine the positions of the knots as a function of time can be more generally and accurately done in the $u-v$ plane, by model-fitting complex interferometric visibilities. This circumvents a number of unwanted errors and allows better accuracy which is otherwise indispensable for detecting motions on kpc-scales, where the time baselines are small and speeds can be low. We use a new generation version of the visibility analysis software DIFMAP \citep{shep94, royc_evn23, royc23_arxiv_ngdif}, that allows modelfitting of "calibration-independent" interferometric closure quantities, in addition to providing robust methods to determine parameter uncertainties.

Table \ref{table:bestpar} shows the best-fit model components and their corresponding parameters (flux density, position, FWHM and position angle), that form the most appropriate description of the source structure. An example of the Gaussian fits can be seen in Figure \ref{fig:vla19}. Note that the jet structure was mostly straightforward, as observed in Figure \ref{fig:vla19}, unlike more complex jets like M87, where simple Gaussians would introduce heavy biases. In this section, we shall discuss the nature of the mean best-fit parameters and how they differ over epoch, in addition to finalizing the positions of the knots and their variances, per epoch.

As visible in Table \ref{table:bestpar}, the 1985 and 2000 epochs were of a slightly worse resolution that the 22 GHz 2003/2019 datasets. Knot A, which is otherwise a point source (the fit axial ratio is $\sim0$), is not apparent in the earliest and lowest resolution datasets, both in the image deconvolution process and in model-fitting. Through several further tests (see e.g., \cite{royc_evn23, royc23_arxiv_ngdif} for detail), we found that the bias introduced in fitting by not including this component is negligible. The claim that the \textit{same} components have been tracked over time can be verified by looking at the broad similarity of component properties like the positions and sizes of the compact Gaussians, with slightly larger sizes (artificial effect) for the lower frequency data due to model-fitting in the absence of larger baselines. For the outer jet (labelled as diffuse emission in Figure \ref{fig:vla19}) instead, modelled by a Gaussian, the positions are at a more significant variance $\lesssim50$ mas or $\lesssim25$ pc, which is allowed given the large size of the component. All of the parameters produced in Table \ref{table:bestpar} are, however, only the best-fit mean values and the extent of their bias or variance is a priori very unclear. Radio interferometric datasets are frequently degraded by lack of $u-v$ coverage and various antenna gain errors, which may create large variances as well as biases in determining the mean value themselves. We follow the procedure of \cite{royc_evn23, royc23_arxiv_ngdif} to estimate the uncertainties. We use the starting basic model for each epoch, and add random amplitude and phase errors and check to what extent the "scatter" in the visibility amplitude and phase "match" that of the real dataset. This gives us an idea of the magnitude of the errors in the respective dataset and is listed in Table \ref{table:gainerr}. Once we had this knowledge of the gain errors, it was then straightforward to run Monte Carlo simulations on the model dataset to see the bias and variance introduced in the parameters by the gain error, where we are mainly interested in the positions of the components. It is natural to see that the bright compact core will be least affected by gain errors.  We found that all datasets except 2003 had negligible bias in the determination of the mean value of the position. The 2003 dataset showed a very strong bias of $\sim+2-3$ mas for X and Y positions of Knot B, and therefore the mean values of the Knot B positions had to be adjusted. The corresponding variance can be obtained from the covariance matrix of the parameter space. However, the same variance is not the correct variance that must be reported in this case. The method implicitly assumes that the observed mean value is the \textit{true} mean value $\mu$ of the parameter histogram obtained after Monte Carlo simulations. This assumption is not always true, which depends on the \textit{probability of the observed value being equal to the mean}. This implies that the observed value is $X_0=(\mu\pm\sigma)\pm\sigma$, or $X_0=\mu\pm\sigma\sqrt{2}$, using a design matrix approach, and where $\mu$ and $\sigma$ are from the Monte Carlo histogram.

We note that in Table \ref{table:bestpar} one finds some discrepancy between the position angles of the Gaussians through various epochs and on an average discrepant sizes within $\sim10$ mas. While this may be because of different frequency observations as noted, we emphasize that this will not affect the determination of the position from the fact that they are \textit{negligibly correlated}, where the correlation is given by the non-diagonal elements of the covariance matrix. Even after the discrepancies, we find that the P.A. of Knot C differs by $\sim10-15$ degrees compared to Knot B. It is possible it represents a change in jet direction and hence has given rise to a stationary shock or substructure, which have been explored in Section \ref{sec:velocity} and Appendix A.

The relevant knot position is only that \textit{relative to the core}. The final \textit{relative} X and Y positions have been tabulated in Table \ref{table:pospar}, along with the corresponding positions \textit{along the jet} and \textit{perpendicular to the jet}, where the reference jet axis is that of the 1985 epoch (obtained by drawing a straight line connecting the 1985 core with the corresponding Gaussian). The errors on the longitudinal and transverse distances have been determined using standard error propagation.

\section{Results and Discussion}
\label{sec:disc}
\subsection{Final Results}

\begin{table*}
\caption{\label{table:speeds} The best-fit apparent speeds for Knots B and C, obtained from a linear least-squares fit The speeds for C, since more complex, have been given in for a pair of epoch ranges. L and T denote longitudinal and transverse respectively. A comma denotes multiple linear fits, as in the case of Knot C. $\bar{d}$ denotes the approximate average projected distance of the knot from the core. The VLBI speed from  \protect\cite{lister19} has been included for completion.}
\begin{tabular}{ccccccc} \hline\hline
Knot & $\bar{d}$ & $\mu_{L}$ & $\beta_{\rm app, L}$ & $\mu_{T}$ & $\beta_{\rm app, T}$\\
         & (pc) & (mas yr$^{-1}$) & & (mas yr$^{-1}$) & \\ 
\hline
VLBI & 1 & & $0.10\pm0.01$ & & & \\
B & 189 & 0.26$\pm$0.07 & 0.51$\pm$0.14 & $0.02\pm0.01$ & $0.04\pm0.02$ \\
C, 1985 to 2003 (T) or 2003 (L) & 335 & $-1.34\pm0.10$ & $-2.60\pm0.20$ & $0.42\pm0.08$ & $0.80\pm0.20$ \\
C, 2000 (T) or 2000 (L) to 2019 & 335 & $0.23\pm0.10$ & $0.50\pm0.20$ & $-0.11\pm0.05$ &  $-0.20\pm0.10$\\
\hline
\end{tabular}
\end{table*}

\begin{figure*}
    \centering
    \includegraphics[width=\linewidth]{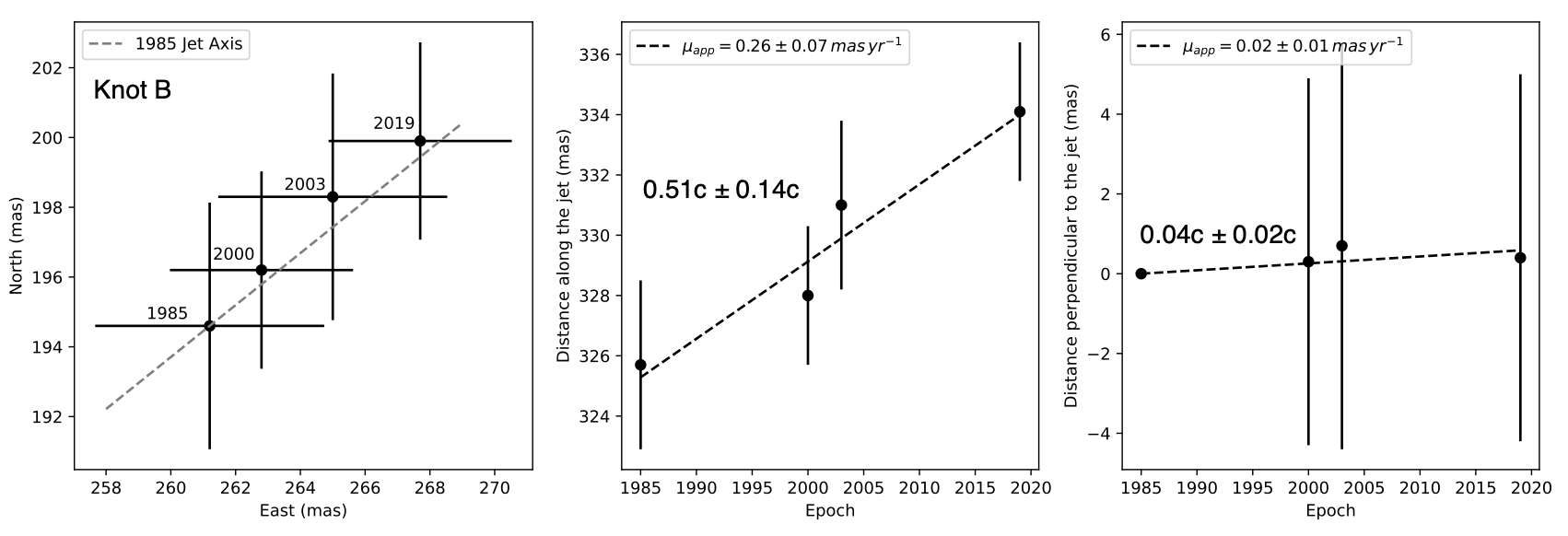}
    \caption{The positional evolution of Knot B with time, from 1985 to 2019. Left: evolution of X and Y positions with time, with the dotted line denoting the jet axis of the 1985 epoch. Middle: evolution of the longitudinal position with time. The linear fit denotes a mildly relativistic speed $\sim0.51c\pm0.14c$. Right: evolution of the position of Knot B in a direction transverse to the 1985 jet axis.  The 1985 position is the reference and hence no error bars can be applied in principle. There is no hint of directed motion, and the speed $0.04c\pm0.02c$ from a linear fit is essentially consistent with zero.}
    \label{fig:knotA}
\end{figure*}

\begin{figure*}
    \centering
    \includegraphics[width=\linewidth]{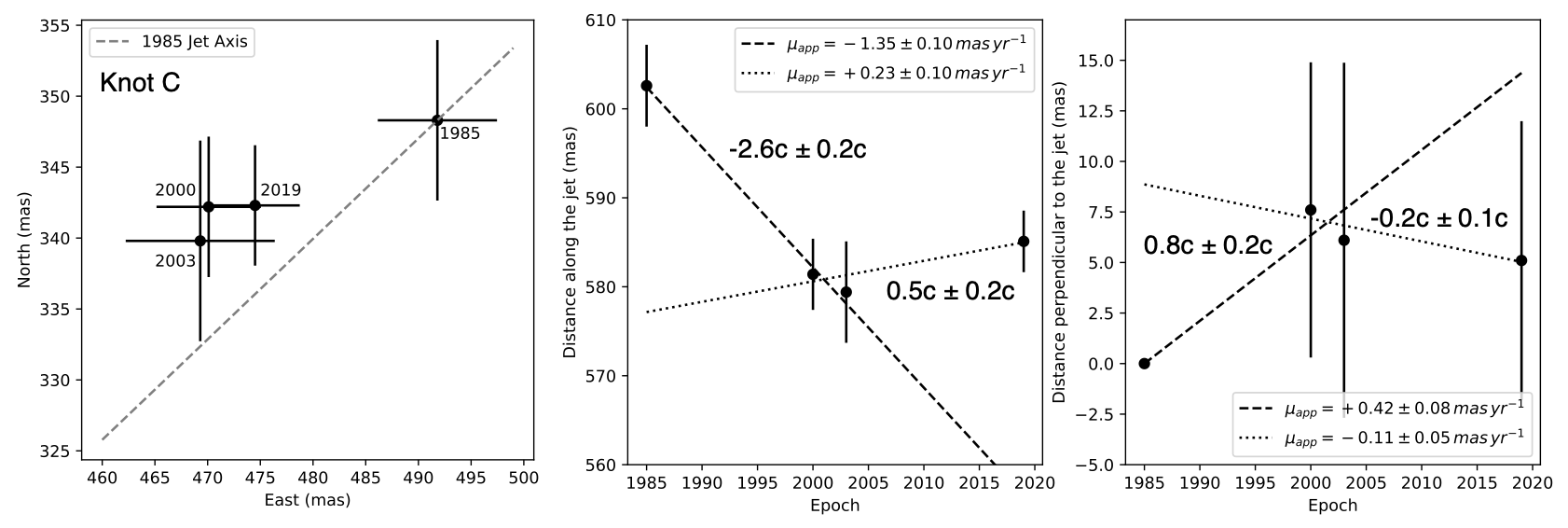}
    \caption{The positional evolution of Knot C with time, from 1985 to 2019. Left: evolution of X and Y positions with time, with the dotted line denoting the jet axis of the 1985 epoch. Middle: evolution of the longitudinal position with time. The linear fit denotes an initial superluminal speed $\sim2.60c\pm0.20c$, but in the backward direction, between 1985 and 2003. Between 2000 and 2019, a linear fit results in a sub-luminal speed of $0.5c\pm0.2c$. Right: evolution of the position of Knot C in a direction transverse to the 1985 jet axis. The 1985 position is the reference and hence no error bars can be applied in principle. There is no indication of systematic motion, and the error bars are large enough for the motion to be consistent with zero.}
    \label{fig:knotB}
\end{figure*}

The positional evolution and the measured speeds for Knot B and Knot C have been shown in Figures \ref{fig:knotA} and \ref{fig:knotB} and tabulated in Table \ref{table:speeds}, using 0.58 pc/mas at $z=0.029$ where $H_0\sim70$ km s$^{-1}$ Mpc$^{-1}$. Knot B shows significant longitudinal motion, at $\beta_{\rm app}\sim0.51\pm0.14$, while it is mostly consistent with zero in the transverse direction. While error bars exist for the positions, they could be not provided for the first epoch of transverse motion since the reference jet axis is chosen based on that, where there cannot be any uncertainty. 

Knot C undergoes apparent superluminal inward motion towards the core between 1985 and 2003 at $\beta_{\rm app}\sim-2.60\pm0.10$, before moving outward at $\sim0.5\pm0.2$ between 2000 and 2019. The transverse motion is similarly odd, with a high apparent speed between 1985 and 2000, after which it is essentially consistent with zero. Note that when only two points are used, the error bar cannot be determined. For the transverse case of the motion between 1985 and 2019 for Knot C, no error bar could be assigned to the 1985 epoch since it is the reference. A superluminal backward motion of a very similar magnitude $\sim2.5$ has also been observed for HST-I in M87 (Harvey/Meyer et al.). 

\begin{figure}
    \centering
    \includegraphics[width=0.9\linewidth]{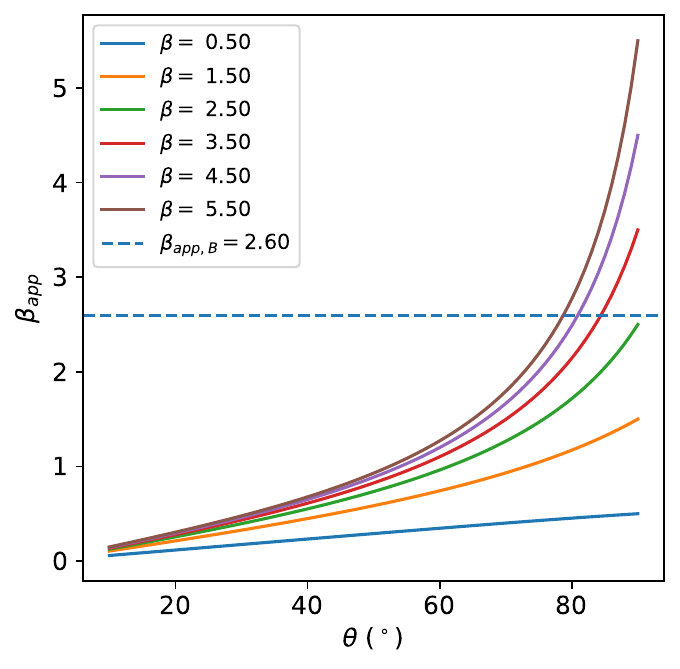}
    \caption{$\beta_{\rm app}$ for inward motion \textit{away} from the observer, for different values of $\beta$. The dotted line is that for Knot C, showing $\theta_{\rm min}\gtrsim70^\circ$ for observing the same.}
    \label{fig:bappin}
\end{figure}

The apparent speed is generally given by $\beta_{\rm app}=\beta\sin(\theta)/(1-\beta\cos(\theta))$, where $\theta\in[0,\pi]$, $\beta,\beta_{\rm app}>0$ (magnitudes) and hence for $\theta>\pi/2$, $\beta_{\rm app}<\beta$. Choosing $\theta\to\pi-\theta$ (since motion is \textit{away} from us) it is evident that the maximum value of $\beta_{\rm app}$ is at $\theta=\pi/2$. Figure \ref{fig:bappin} shows $\beta_{\rm app}$ for various values of $\beta$, with the corresponding $\beta_{\rm app}$ for Knot C given in a dotted line. While it is clear that $\beta>1$ for $\theta=\pi/2$ to produce $\beta_{\rm app}>1$ at $\theta<\pi/2$, for low viewing angles $\theta<\pi/3$, in spite of how large $\beta$ is, $\beta_{\rm app}<1$. This implies Knot C \textit{must} have $\beta>1$ at $\theta=\pi/2$ and must be viewed at $\theta\gtrsim\pi/3$ to produce the observed superluminal motion, as also observed in Figure \ref{fig:bappin}. This is at variance with the expectation of the viewing angle of this source, which is $\sim\pi/9$ ($20^\circ$), expected from the absence of the counterjet and also beamed emission: see Section \ref{sec:sed} for the spectral energy distribution (SED) modelling of the extended jet, where we have approximately constrained $\theta\sim20^\circ$. Furthermore, since it is clear that a single \textit{physical} component cannot have $\beta>1$, Knot C must be an \textit{unresolved} knot containing substructure where plasmoids move independently to cause backward superluminal motion or where the magnetic field structure peculiarly depends on distance $z$ along the jet. We have further discussed this and a possible reconciliation of low viewing angle with that required by Knot C, using theoretical modelling in the next section.

\subsection{Hydrodynamic Jet Velocity, Velocity of Moving Knots}

\label{sec:velocity}

A bright moving pattern in the jet can represent a host of phenomena (see e.g., \citealt{phinney85}). Since knots are thought to be regions in the jet where particle acceleration occurs, the idea of a shock front energizing particles has been frequently adopted in many works. Shocks may arise due to pressure mismatch at the jet boundary or due to sudden differences in jet properties and are routinely observed in hydrodynamic simulations of relativistic jets. However, it is a priori unclear how this shock velocity may relate to the velocities of the upstream and downstream plasma. Given shock initial conditions, say $\bm{\beta'_{\rm shock}}$ is the shock velocity in the frame of the pre-shock (upstream) fluid moving at $\bm{\beta_{\rm pre}}$, the shock boundary conditions can be numerically solved to obtain an estimate of the post-shock (downstream) fluid velocity $\bm{\beta'_{\rm post}}$ in the frame of the pre-shock fluid and thereafter in the lab frame. The corresponding methods have been very clearly discussed in, for example, \cite{bland76,konig80,bb96}. 

However, this is only in the context of fluid dynamics and these speeds cannot be "observed". Incorporation of particle acceleration due to the shock readily allows to "track" the motion of that part of the post-shock fluid that is the brightest. In the simplest case of a non-decaying shock moving through the jet, the brightest emission inevitably arises from the fluid just downstream of the shock, assuming the shock is infinitesimally thin. Therefore in this model, the observed "pattern" speed along the direction of jet flow $\hat{z}$ will essentially reflect the speed of the shock, as in $\bm{\beta_{\rm pattern}}=\beta_{\rm shock}\bm{\hat{z}}=\frac{\beta_{\rm pre}\pm\beta'_{\rm shock}}{1\pm\beta_{\rm pre}\beta'_{\rm shock}}\bm{\hat{z}}$, where the $\pm$ incorporates the direction of the $\bm{\beta'_{\rm shock}}$. This clearly implies that a "bulk" jet velocity can be defined to be either of $\beta_{\rm pre}$ or $\beta_{\rm post}$ in a region containing a shock front moving with a pattern speed $\beta_{\rm shock}$. For $\bm{\beta'_{\rm shock}}\cdot\hat{z}>0$, it is clear that $\beta_{\rm pattern}>\beta_{\rm pre},\beta'_{\rm shock}$, implying the speed of a knot is only a loose upper limit to pre-shock bulk velocity $\beta_{\rm pre}$ in this simplest model. Since the shock speed must also be supersonic with respect to the pre-shock fluid, large $\beta_{\rm pattern}$ must imply that $\beta_{\rm pre}$, or the pre-shock bulk speed, is more likely to be large. This is in accordance with pc-scale proper motion observations, where superluminal pattern speeds directly imply that the bulk speed is also likely to be relativistic. For extremely high $\Gamma_{\rm shock}$, $\Gamma_{\rm pre}\sim\Gamma_{\rm shock}$, where we must note that for $\Gamma\gtrsim7$, a $\Delta\Gamma=\Gamma_{\rm pre}-\Gamma_{\rm shock}\sim1$ implies $\beta_{\rm shock},\beta_{\rm pre}\gtrsim0.99$ or so. However, when $\Gamma\lesssim2$, the fluid is not highly relativistic and a $\Delta\Gamma\sim1$ may overlook \textit{very different} plasma and shock velocities. While for VLBI observed jets, superluminal motion is more common, it is mostly immaterial to discuss the plasma and pattern speeds. For the case where $\bm{\beta'_{\rm shock}}\cdot\hat{z}<0$, it is evidently very difficult to constrain a bulk speed without involved modelling. For mildly relativistic $\beta_{\rm app}$ observations ($\Gamma\lesssim2$), which is true for some MOJAVE observations, as well as which is expected for the large-scale jets, the relation between $\beta_{\rm app}$ and $\beta_{\rm pre}$ must therefore be clarified on a case-by-case basis. One possible solution is observing motions of multiple knots and looking for systematic similarities. However, no tighter constraint can be derived on $\beta_{\rm pre}$ from the speed measurement of a single knot when it is not ultra relativistic, like all the measured speeds of this source, shown in Table \ref{table:speeds}. In this sub-section we therefore discuss the case of Knots B and C, and related theoretical modelling to infer the bulk flow properties of the large-scale jet.

\subsubsection{Case of the VLBI speeds and Knot B}

VLBI observations of pc-scale jets often show ejections of bright components from the unresolved core, that propagate outward along the jet at \textit{constant} velocities. Such systematic observations imply they are moving outward, in the frame of the pre-shock fluid (or $\bm{\beta'_{\rm shock}}\cdot\bm{\beta_{\rm pre}}>0$). This implies that the maximum observed VLBI speed in 3C 78, which is $\beta_{\rm pattern}\sim0.22$ (at $\theta=20^\circ$ for $\beta_{\rm app}=0.1$), implies that $\beta_{\rm pre}<0.22$. However, an inward moving shock ($\bm{\beta'_{\rm shock}}=-\beta'_{\rm shock}\bm{\hat{z}}$), but with a velocity $\beta'_{\rm shock}<\beta_{\rm pre}$ can result in a slow outward motion of the knot. \cite{lister19} find consistent outward motion for eight components in the VLBI jet of 3C 78. If there were varying signs of $\bm{\beta'_{\rm shock}}$, combinations of forward and backward motion would have been motions. Therefore $\bm{\beta'_{\rm shock}}\cdot\bm{\beta_{\rm pre}}>0$ is more probable. Using a similar argument for Knot B, the constraint we can apply is $\beta_{\rm pre}<0.6$ ($\beta_{\rm pattern}=0.6$ for $\theta=20^\circ$).

\subsubsection{Motion of Knot C}

The case for Knot C is considerably more complex. It first moves inward at a superluminal speed $\sim2.6c$, then moves outward at $\sim0.5c$. Since the peak of the emission is only one point as any substructure can hardly be resolved (further, in a radio interferometric \texttt{clean} image, minute substructure is highly dependent on the model chosen by the user), the apparent speed for a backward moving unresolved emitting point in the jet is simply $-\beta_{\rm app}=-\frac{\beta_{\rm pattern}\sin\theta}{1+\beta_{\rm pattern}\cos\theta}$, where $\beta_{\rm app}<\beta_{\rm pattern}$, as discussed in Section \ref{sec:velocity}. If the jet is fully resolved, this rules out the possibility of a single shock moving superluminally towards the core and opens up multiple possibilities. 

Using an analytical model of a stationary shock and natural toroidal magnetic field configuration, the evolution of the brightness distribution of the post-shocked fluid can be followed. The details of the model and its parameters are given in Appendix A. In Figure \ref{fig:fitadhoc}, we have attempted to indicatively fit the observed positions of Knot C using our model. Using the transformations in Appendix A, and convolving $j_\nu(z,t)$ with a beam size of desired resolution, we plot the evolution of the peak emissivity position with time and its velocity. For our purpose, we use a beam size of width $0.1"$, corresponding to the VLA K-Band A-config (22 GHz) resolution. We used $\beta_{\rm post}=0.7$, $\alpha=0.5$, $\theta=75^\circ$, $m=6$ and $z_{s0}=90$ pc, where $\beta_{\rm post}$ refers to the speed of the post-shocked fluid in the frame of the pre-shocked fluid, $\alpha$ is the synchrotron spectral index across the shock, $\theta$ is the viewing angle, $z_{s0}$ is the position of the stationary shock and $m$ is a numerical parameter governing the symmetry of magnetic field lines around the shock. While the superluminal backward speed could be obtained self-consistently, we found it difficult to describe the positional evolution from 2003 and 2019 using our simple model. However, a modification of $m$ and $z_s$ allowed a closer fit as evident, except with the presence of a number of harmonics in both position and velocity space, due to $m\neq1$. A scenario like this seems partially improbable since the resulting harmonics are of a lower magnitude and do not produce the observed superluminal motion, unlike a periodic pattern for $m=1$. Future monitoring of 3C 78 is required to constrain the same. 

We have hence approximately established a possible physical scenario of Knot C and its superluminal motion. However, it is unclear if the required high $\theta$ can be by an intrinsic bend in the jet. It may be probable that since at large scales, where $\beta\sim0.5-0.7$, the intrinsic jet opening angle is large $\theta\gtrsim45^\circ$, the observed motion of Knot C is along the outer sheath/surface of the jet, which must make a larger angle to our line of the sight than the jet "spine". 

If we assume that the viewing angles are large especially for Knot C, we must also consider how $\beta_{\rm pre}$ may be related to $\beta_{\rm post}$. For the previous figures, it was clear that a low $\beta_{\rm post}$ can produce the observed superluminal motion. However, a too low $\beta_{\rm post}\lesssim0.3$ will not produce the required motion as well as the velocity magnitudes. Therefore, $\beta_{\rm post}$ must be larger than a minimum value. Further,  $\beta_{\rm pre}>\beta_{\rm post}$ in a shock, or $\beta_{\rm pre}\gtrsim\beta_{\rm post,B}=0.7$, from Figure \ref{fig:fitadhoc}.

However, as discussed above, the model, while producing the overall forward and backward motion of the peak emission position around a stationary shock, suffers from a few shortcomings that include a prediction of harmonics after $\sim450$ years and a specified magnetic field pitch angle model. It is unclear what may cause such a configuration self-consistently if turbulence is present, which is highly likely in extragalactic jets.

The existence of backward and forward motion can also be explained using relativistic gas dynamics. Following \cite{daly88}, in a jet expanding into a constant pressure medium at $P_{\rm ext}$ from $P_0>2P_{\rm ext}$, a shock front forms where two or more characteristic curves interest. As $P_{\rm ext}$ is decreased further, the shock forms at a position further upstream. The dynamics of Knot C can hence be described as follows. At a time $t=t_0$, a shock front is formed at $z=z_0$. Let us assume the total time to produce a fluctuation in $P_{\rm ext}$ at a given spatial region around the jet and thereafter create a shock at $z<z_0$ is $t_{\rm shock}$. If it takes $t_{\rm dec}$ for a shock to lose its strength appreciably, at $t>t_0+t_{\rm dec}$ non-thermal electron injection can be assumed to have stopped. This implies that if $t_{\rm shock}<t_{\rm dec}$, it is evident that two shocks of unequal brightness may exist within an unresolved knot if two or more characteristics intersect within $100$ pc, which is not unlikely. The shock further upstream would be brighter resulting in a sudden backward motion of the knot brightness centroid. If this phenomenon repeats and both of the shocks move downstream while decaying at the same time, this will result in a forward motion following a backward motion, and the repetition of the same phenomenon. This would explain the motion of Knot C. We have only provided a very qualitative explanation as the presence of multiple shock fronts alongside the breakdown of the steady state assumption highly complicate the analysis. This would require a radiation hydrodynamics simulation and is out of the scope of this paper.

A similar pattern of motion has also been observed in HST-I ($\lesssim100$ pc, Meyer \& Harvey, in prep.), which undergoes superluminal backward motion followed by forward motion over a span of $\sim 20$ years. While that will be discussed in a future paper, existence of this similarity hints at possibly similar physical processes occurring at HST-I and Knot C.

\begin{figure*}
    \centering
    \includegraphics[width=\linewidth]{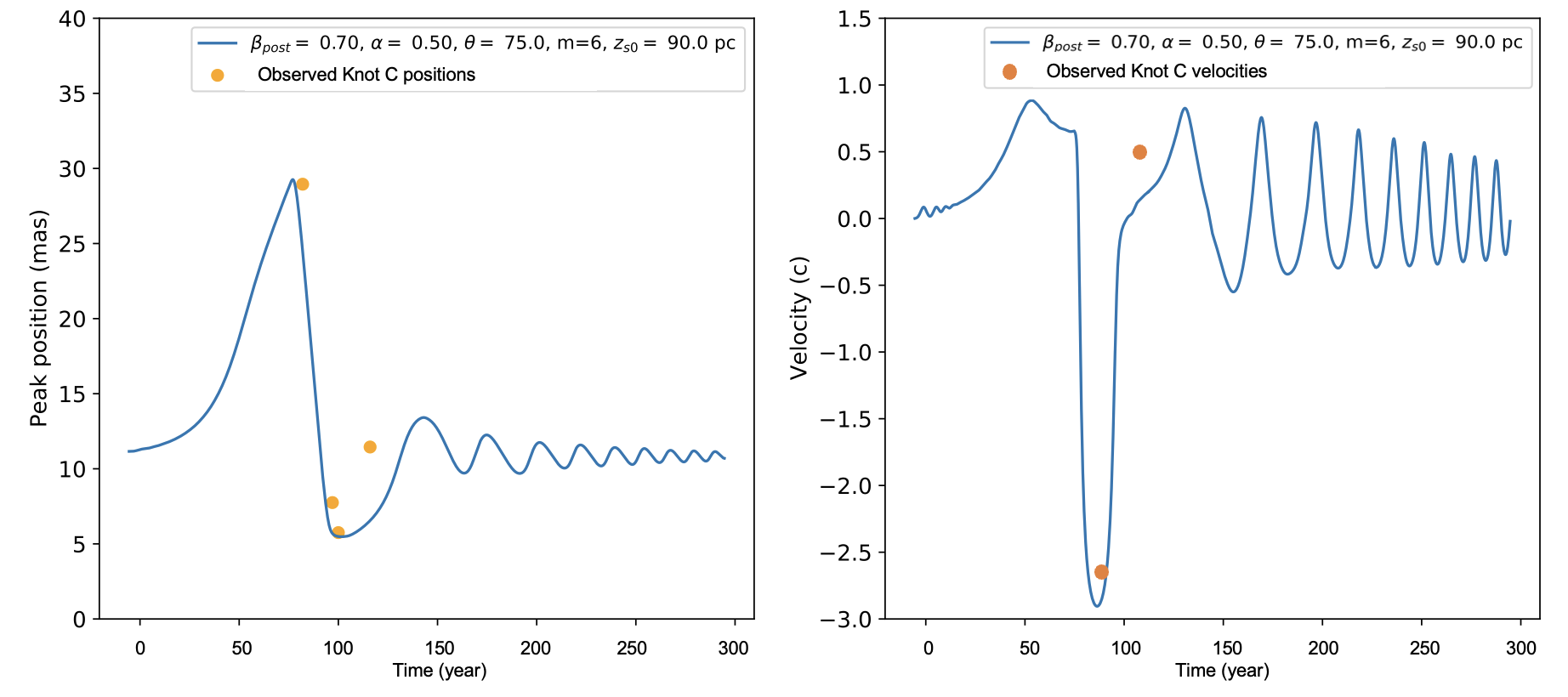}
    \caption{Left: peak position evolution for parameters given in the panel, to fit the observed Knot C speeds, with the starting time being arbitrary and direction being longitudinal. While the superluminal backward motion can be represented by the most simple model, the resulting outward motion could be approximately obtained using ad-hoc values of $m=6$ and $z_s=90$ pc, where it is unclear if the former is physically plausible. Right: evolution of the apparent velocity with time, with orange points showing the observed velocities. The harmonics are a result of $m>1$.}
    \label{fig:fitadhoc}
\end{figure*}

\subsubsection{Velocity profile and evidence of bulk acceleration}

\begin{figure}
    \centering
    \includegraphics[width=0.9\linewidth]{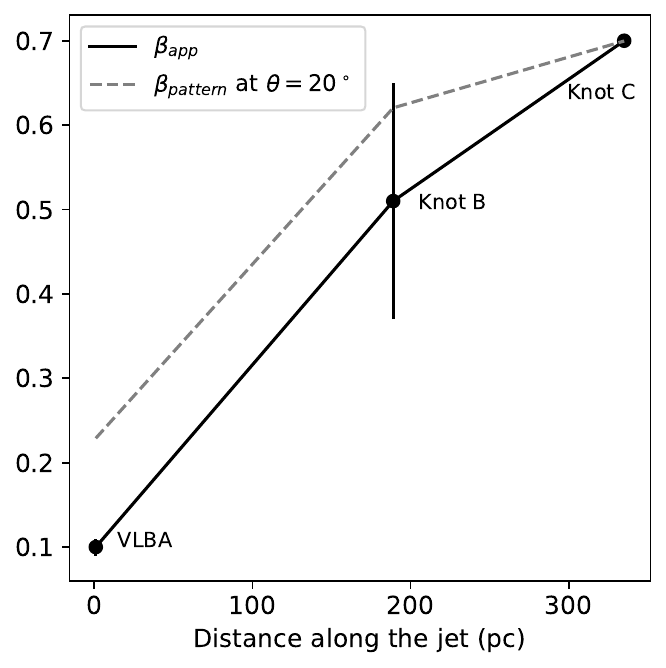}
    \caption{$\beta_{\rm app}$ (solid line) versus projected distance $z$ along the jet, from the central engine. It is evident that $\beta_{\rm pattern}$ (dotted line) is indeed increasing and such a systematic increase can only be explained by an increase in the bulk flow speed with $z$. Therefore, this implies that the jet is either still accelerating at the $\sim$ 100 pc scales, or has accelerated in the meso-scales and is decelerating at the VLA scales.}
    \label{fig:velprofile}
\end{figure}

The observed $\beta_{\rm app}(z)$ has been plotted as a function of $z$ in Figure \ref{fig:velprofile}, where the dotted line denotes the corresponding $\beta_{\rm pattern}$ if $\theta=20^\circ$ is used for the VLBI knot and Knot B, with $\beta_{\rm pattern,C}=\beta_{\rm pre,C}\sim\beta_{\rm app,C}$ from Section 4.2.2 using $\beta'_{\rm shock}=0$ and large viewing angle $\theta_C$.
We find $\beta_{\rm pattern}$ increasing with distance along the jet, $z$, from the central engine. A similar profile would be observed even if one would take $\theta_C=20^\circ$ and $\beta_{\rm app,C}=0.5$. While a non-increasing $\beta_{\rm pre}(z)$ cannot be ruled out if $\beta'_{\rm shock}(z)$ is systematically increasing with $z$, the environmental conditions in which a jet propagates change strongly through parsec to kilo-parsec scales and it is difficult to predict any general property of $\beta'_{\rm shock}(z)$. Therefore, $\beta'_{\rm shock}(z)$ cannot monotonically increase with $z$ and an increasing $\beta_{\rm pattern}(z)$ can only imply that $\beta_{\rm pre}(z)$ is also increasing with $z$, or the jet is undergoing bulk acceleration between the parsec and hundred parsec scales.

\subsection{Jet profile from pc to kpc-scales}
\label{sec:profile}
\begin{figure}
    \centering
    \includegraphics[width=\linewidth]{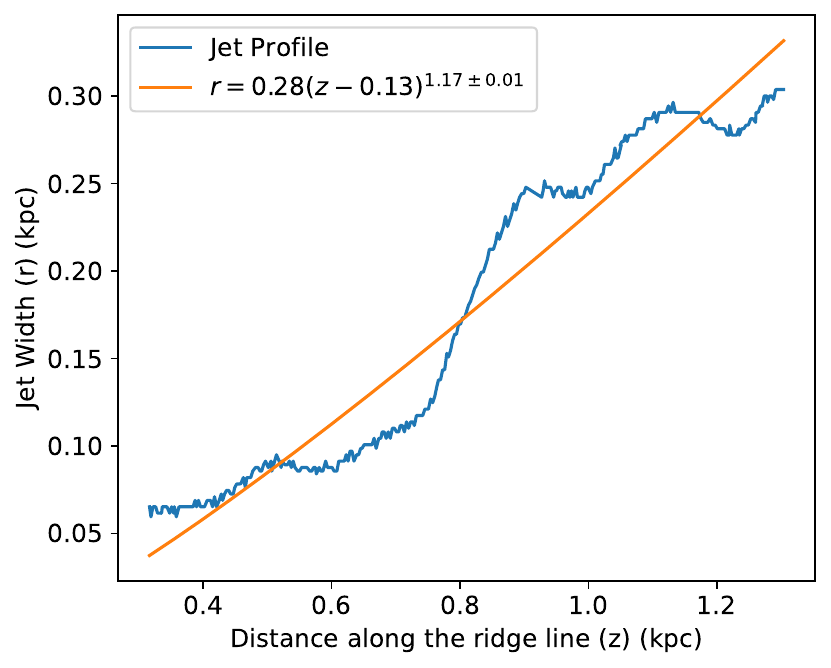}
    \caption{The jet width $r$ v/s projected distance $z$ along the jet/ridge line, between which there is an approximate linear relationship, $r\sim z^{1.17\pm0.01}$, implying the jet shape is approximately conical.}
    \label{fig:profile}
\end{figure}

In this sub-section we discuss the jet shape from parsec to kilo-parsec scales. For the parsec scales, we find $r\propto z$ from \cite{pushkarev17}, where $r$ is the jet width at distance $z$ along the jet from the central engine. The jet profile is calculated as follows for the VLA observation, where stacking multiple images is irrelevant due to the overall constancy of observed structure over a 40 years. To calculate the profile, we chose the 2019 22 GHz observation (Figure \ref{fig:vla19}) that has the best quality and resolution. We determine a "ridge line" along the jet, which is the position of maximum intensity in the jet surface at a given distance along the jet. At every ridge point, \cite{pushkarev17} fits a Gaussian across the jet width and uses the corresponding FWHM as a measure of the width $r$. However, prescription of any general method for the same is ill-defined due to the ill-defined nature of the jet "boundary". Further, a major assumption here is that the jet brightness cross section does not strongly vary across $z$, which breaks down at knots. In this case, we use the same prescription for easy comparison and plot the profile in Figure \ref{fig:profile} as a function of $z$ along the ridge-line. We choose the minimum $z$ just beyond the position of knot C ($\sim 350$ pc) to ensure bias due to presence of steep brightness profiles like knots is reduced. We find, using a power law fit of $r=a(z+z_0)^k$, that $k\sim1.17\pm0.01$, where the slope is higher than 0.5 (which is expected for a parabolic profile) and closer to 1, expected for a conical profile. Therefore we find that the jet of 3C 78 remains approximately conical from parsec to kilo-parsec scales.

\subsection{SED modelling of the extended jet}
\label{sec:sed}
\begin{figure*}
    \centering
    \includegraphics[width=0.8\linewidth]{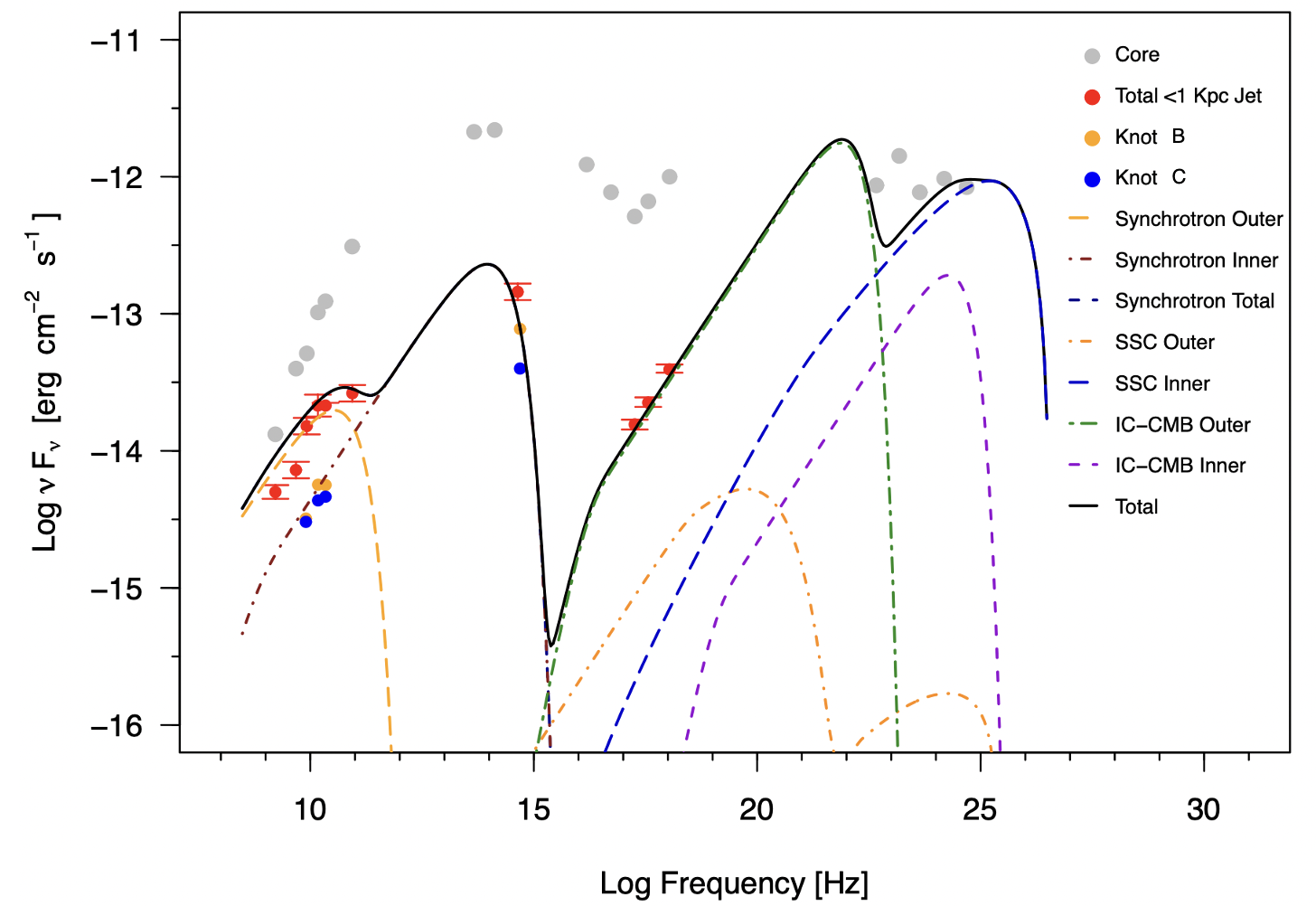}
    \caption{Observed and modelled spectral energy distribution of 3C 78. While the gray points show unresolved/core emission, the red points are fluxes from the total arc-second scale $<1$ kpc jet. The orange and blue points are that of Knots B and C. The total SED model is given a black solid line, while the sub-components have been marked with different dashed lines of differing colours. The synchrotron, SSC and IC/CMB models of the outer zone are given in orange dashed, orange dot-dashed and green dot-dashed respectively. For the inner zone, the synchrotron, SSC and IC/CMB models are given in brown dot-dashed, blue dashed and violet dashed respectively.}
    \label{fig:sed}
\end{figure*}

In this section we discuss the SED of the extended jet of 3C 78, using multi-wavelength observations. This is tantamount to our understanding of jet acceleration and emission models for this specific source. The spectral energy distribution (SED) of the extended (HST/VLA) jet has been plotted in Figure \ref{fig:sed}. The core/unresolved points have been given in grey, while the total $<1$ kpc jet fluxes are given in red. The jet synchrotron spectrum is unusually broad, ranging through nearly 6 decades in frequency. The fact that it is not an artifact of using the "total" jet flux (thereby ignoring inhomogeneities like knots), can be seen in a similar pattern of a knot by knot radio-optical spectra, where the fluxes of Knots B and C are given in orange and blue respectively. In contrast, since the X-ray jet is less resolved than the radio/optical, the X-ray fluxes mainly represent the total extended jet.

From Tables \ref{table:other_fluxes} and \ref{table:vla}, it is evident that neither the unresolved nor the jet fluxes are simultaneous. core data points are not simultaneous. Since it is true that source variability can produce changes that are not captured by our observed spectral energy distribution, only an indicative SED fit could be made. However, we note that \cite{fukazawa15} find no flares in the five-year Fermi light curve. Due to the complexity of the SED, we used a superposition of two one-zone leptonic models to produce a fit to the entire SED of the extended ($<1$ kpc) jet of 3C 78. Our code models both synchrotron and inverse Compton emission (external Compton and Synchrotron self-Compton) with Klein-Nishina effects. For each zone, we assume the emitting region has size $R$ and is filled with a uniform magnetic field $B$. It moves with a bulk Lorentz factor $\Gamma$ at an angle $\theta$ to our line of sight (the Doppler factor $\delta$ can be inferred). Electrons of energy $\gamma$ are injected initially using the following power-law prescription $Q(\gamma,t)=N_0\gamma^{-s}\delta(t)$. The electron energy distribution $n(\gamma,t)$ is hence evaluated by solving the time-dependent continuity equation:

\begin{equation}
\frac{\partial n(\gamma,t)}{\partial t}+\frac{n(\gamma,t)}{t_{\rm esc}}=\frac{\partial}{\partial \gamma}[\dot{\gamma}n(\gamma,t)]+Q(\gamma,t)
\end{equation}

where $t_{\rm esc}$ is the average time an electron spends in the emission region and the first and second terms on the right hand side represent cooling and injection respectively.

Our two-zone SED consists of two sets of physically motivated parameters belonging to a heated shocked plasma and the shocked plasma that has expanded thereafter and partially cooled due to partial adiabatic, synchrotron and Compton losses. A few properties of the expanded plasma can be directly inferred from the initial parameters. For example, the electron luminosity, the magnetic field and $\gamma_{\rm min}$, $\gamma_{\rm max}$ must decrease with expansion. However, modelling of the same for anisotropic emission requires a multi-zone approach and complicated time-dependent calculations. This can be approximately circumvented in the following way. The "inner" plasma (or the one recently shocked) must radiatively cool only by the highest energy electrons and hence $\gamma_{\rm min,in}\gg\gamma_{\rm min,0}$ and $\gamma_{\rm max,in}=\gamma_{\rm max,0}$. Since the "outer" plasma is a result of incomplete cooling of the inner plasma, it can described separately with $\gamma_{\rm min,out}=\gamma_{\rm min,0}$ and $\gamma_{\rm max,out}\ll\gamma_{\rm max,0}$, but with a lower electron luminosity, taking into account adiabatic expansion. The same allows a larger size, lower magnetic field and larger electron escape time-scale $t_{\rm esc,out}$. Simple Bohm diffusion arguments for cosmic ray electrons \citep{ohira12} also predict larger $t_{\rm esc}$ for lower energy electrons. 

The best-fit SED model has been plotted in Figure \ref{fig:sed}. The black solid line represents the total modelled spectrum. The synchrotron, SSC and IC/CMB models of the outer zone are given in orange dashed, orange dot-dashed and green dot-dashed respectively. For the inner zone, the synchrotron, SSC and IC/CMB models are given in brown dot-dashed, blue dashed and violet dashed respectively. The total spectrum can be evidently well described by using this two-zone model. The relevant physical parameters of these two zones have been tabulated in Table \ref{tab:par} and are physically viable given our initial prescription. 

\begin{table}
\centering
\caption{Physical parameters of the SED model for the core and the extended jet emission. $R_{\rm em}$ refers to the size of the emission region and $L_e$ is the injected electron power.}
\label{tab:par}
\begin{tabular}{lll}\hline
Parameter & Inner Zone & Outer Zone \\
\hline
$\delta$ (Doppler factor) & 2.4 & 3 \\
$\Gamma$ (Bulk Lorentz factor) & 4 & 4 \\
$\theta$ & $18.9^{\circ}$ & $22.3^{\circ}$ \\
$R_{\rm em}$ (pc) & 23 & 166 \\
$t_{\rm esc}$ ($R_{\rm em}$/c) & 1 & 12  \\
$s$ & 2 & 2 \\
$\gamma_{\rm min}$ & $2\times10^{3}$ & 60  \\
$\gamma_{\rm max}$ & $1.4\times10^6$ & $7\times10^4$ \\
B (G) & $1.23\times10^{-5}$ & $1.5\times10^{-6}$ \\
$L_e$ (erg $s^{-1}$) & $1\times10^{45}$ & $6\times10^{44}$ \\
\hline
\end{tabular}
\end{table}

We find that the above model can quite sufficiently reproduce the entire nature of the total $<1$ kpc jet spectrum. The broad synchrotron spectrum can be described well as a sum of inner and outer synchrotron emission models. The optical flux is under-predicted by $\sim20\%$, which may occur due to the approximate nature of the model. The jet X-rays are found to match an IC/CMB model, that predicts a high value of $\lesssim$ MeV to tens of MeV flux. However, since we do not have observations in that range, we cannot test that yet. In contrast, SSC in the inner zone explains the $\sim10$ GeV flux and predicts significant emission around $\sim100$ GeV. The SSC is considerably dormant in the outer zone mostly due to the much lower magnetic field and electron luminosity. We note that for electrons in the outer zone, the synchrotron energy density of photons from the inner zone must be very close to the synchrotron energy density in the outer zone itself, since the latter is an \textit{effect} of the former due to adiabatic expansion. For electrons in the inner zone, the energy density of the synchrotron photons from the outer zone is obviously much lower, around $\ll(B_{\rm out}/B_{\rm in})^2\simeq0.01$ times less than the synchrotron energy density in their own zone, using Table \ref{tab:par} and also drop of photon energy density with distance. Therefore, we have neglected inter-zonal external Comptonization of synchrotron photons (EC/Synch). Note that this approximation is only valid since we have large size ratios at hand, as in $R_{\rm in}\sim25$ pc and $R_{\rm out}/R_{\rm in}\sim5$. 

In contrast, for a two-zone model not motivated by physical considerations, we have found that using electron luminosities $\sim$ 100 times higher for electrons having lower energies than the putative inner zone, the X-rays can be explained using an SSC model. We have discarded it on the lack of physical grounds. However, observations of the source in hard X-rays will clearly be able to test the origin of the soft X-rays, where the two competing two-zone models can be tried. We also note that the core synchrotron spectrum is also visibly broad and most likely requires multi-zone modelling. Tests of the same using variability can also be made by monitoring the source at multiple wavelengths.

Our knowledge of the basic character of the kpc-jet of 3C 78 can be used to derive basic estimates of jet magnetization at these spatial scales. For the shocked plasma that is actually responsible for the entire observed SED, magnetic fields and electron luminosities will obviously be higher than that in a cold plasma. For the maximum estimate of the magnetization $\sigma=P_B/P_e$, we can compute the magnetic field power of the outer zone and compare it with the jet kinetic power of a purely leptonic $e^{-}-e^{+}$ plasma that has adiabatically expanded to ten times the initial size (like the outer zone). If we assume an acceleration mechanism only energizes fraction $\eta_e=1\%$ of the electrons, the total power in the leptons would then be $P_{\rm e^-e^+}=P_{\rm cold}+P_{\rm hot}=\frac{1-\eta_e}{\gamma_{\rm min}\eta_e}P_e+P_e\simeq2.7\times10^{45}$erg/s for $\langle\gamma\rangle\simeq\gamma_{\rm min}=\gamma_{\rm min,out}=60$. The corresponding shocked magnetic field power (using $R_{\rm out}$ and $B_{\rm in}$ for conservative estimate) would be $P_B=\pi R_{\rm out}^2\Gamma^2 U_{\rm B,in}\beta c\simeq2\times10^{42}$erg/s (for $\beta\simeq1$). This provides us an estimate for the magnetization $\sigma=P_B/P_{\rm e^-e^+}\simeq10^{-3}\ll1$, which must be even lower for a hadronic jet. This is expected at such large distances away from the central engine. Therefore the $\sim1$ kpc jet of 3C 78 can be considered to be matter-dominated.

\subsection{Bulk acceleration in a matter-dominated jet}

The prevailing "standard model" of magnetic acceleration refers to the continual conversion of magnetic energy into directed bulk kinetic energy as the jet expands sideways \citep{komiss09}. This is generally expected to occur as long as the jet is magnetically dominated, where it reaches a terminal Lorentz factor mainly governed by the initial magnetic energy density and a "bunching" factor of the magnetic field lines \citep{komiss09}. Recent work by \cite{kov20} on parsec-scale jet profiles of a large sample of jets from AGN hints at a possible change of jet profile from parabolic to conical coincident with a transition from a magnetic to matter-dominated regime, where the jet has reached the asymptotic Lorentz factor. Indeed, the jet of M87 displays similar behaviour, where the jet profile after HST-I, where it starts decelerating, changes from a parabolic to conical shape. However, we must note that the theory does not very specifically predict that large magnetic energy densities must result in more "bound" or parabolic profiles. 

Our observations for the jet of 3C 78 are in stark contrast to the above. While we find that the jet has accelerated or is still accelerating between the parsec and hundred parsec scales, the jet profile is highly conical from parsec to kilo-parsec scales, where the hundred parsec-scale jet is strongly matter-dominated. While at a certain level (here less than the VLBI scales), the initial acceleration of the jet \textit{must} be magnetic, which occurs in the acceleration and collimation zone, it is likely that magnetic fields are not responsible for producing the \textit{observed} bulk acceleration of the jet of 3C 78. The other possibility is that of acceleration by simple conversion of random internal energy of molecules into directed kinetic energy ("thermal acceleration"), as would occur in a matter-dominated hydrodynamic jet that starts out as barely supersonic with respect to the ambient medium and expands with a given equation of state. This is similar to the de Laval nozzle discussed in \cite{bland78}, where due to very carefully assigned pressure boundary conditions before and after a nozzle, one may continually accelerate bulk from a subsonic to a supersonic speed through the nozzle. The situation can be described more specifically through straightforward one-dimensional Euler equations, for the adiabatic and the "quasi-isothermal" case, discussed by \cite{crumley17} in a paper that revises some of the results of \cite{falcke95} and their definition of a "maximal jet". In the adiabatic case, the internal energy is continually used to accelerate the jet as it expands and the jet reaches the "terminal" Lorentz factor $\Gamma=\Gamma_0(1+\gamma\zeta)$ (where $\zeta$ is the fraction of the rest mass energy that equals the initial internal energy, $\gamma$ is the adiabatic index and $\Gamma_0$ is the initial Lorentz factor of the flow), corresponding to the case when all internal energy has been used up. This is in contrast to the quasi-isothermal case, where the temperature, although does not remain constant, goes as $T\propto(\Gamma\beta)^{1-\gamma}$ and the internal energy falls off slower with $z$ than the adiabatic case. This allows a terminal Lorentz factor larger than that allowed through energy conservation, since the jet is assumed to be heated \textit{continually}. However, as \cite{crumley17} mention, it is unclear what may allow a continual heating of the jet plasma to prevent adiabatic losses. Magnetic reconnection at hundred parsec scales is not preferred due to matter-dominance. It must be noted that a thermal acceleration model already invokes the concept of a "temperature", that presupposes the notion of a thermal equilibrium. However, it is clear that the jet of 3C 78 possess strong shocks containing compressed magnetic fields, where the particle spectra is non-thermal. This attempt, therefore, is only to characterize the possibility of an alternative to magnetic acceleration models in the large-scales. Additionally, a smooth bulk Lorentz factor profile in the presence of knots is impossible. While a number of open questions regarding thermal bulk acceleration of jets remain, in this section we aim to understand the parameter requirements in the adiabatic and quasi-isothermal models to produce the order of the acceleration observed in this case.

Using \cite{falcke95} or \cite{crumley17}, we numerically solve the one-dimensional Euler equations for $\gamma=5/3$ (non-relativistic) for a conical adiabatic and quasi-isothermal jet, for different values of $\zeta$. The adiabatic Euler equation, where the temperature $T\propto(\Gamma\beta)^{1-\gamma}z^{2-2\gamma},$ is given by:

\begin{equation}
\begin{split}
\bigg[\Gamma\beta\frac{\gamma+\xi}{\gamma-1}-\gamma\Gamma\beta-\gamma/(\Gamma\beta)\bigg]\frac{d(\Gamma\beta)}{dz}=\frac{2\gamma}{z}(1+\Gamma^2\beta^2),\\
\xi=(\Gamma\beta/\Gamma_0\beta_0)^{\gamma-1}(z/z_0)^{2\Gamma-2}/\zeta
\end{split}
\end{equation}

where $\Gamma_0\beta_0$ refers to the four-velocity of sound at the sonic point $z_0$ where the jet is launched, or $\Gamma_0\beta_0=\sqrt{[\zeta\Gamma(\Gamma-1)][1+2\zeta\Gamma-\zeta\Gamma^2]^{-1}}$.

For the quasi-isothermal case, the equations are slightly different since the temperature does not fall off with $z$ explicitly:

\begin{equation}
\begin{split}
\bigg[\Gamma\beta\frac{\gamma+\xi}{\gamma-1}-\gamma\Gamma\beta-\gamma/(\Gamma\beta)\bigg]\frac{d(\Gamma\beta)}{dz}=\frac{2}{z},\\
\xi=(\Gamma\beta/\Gamma_0\beta_0)^{\gamma-1}/\zeta
\end{split}
\end{equation}

\begin{figure}
    \centering
    \includegraphics[width=\linewidth]{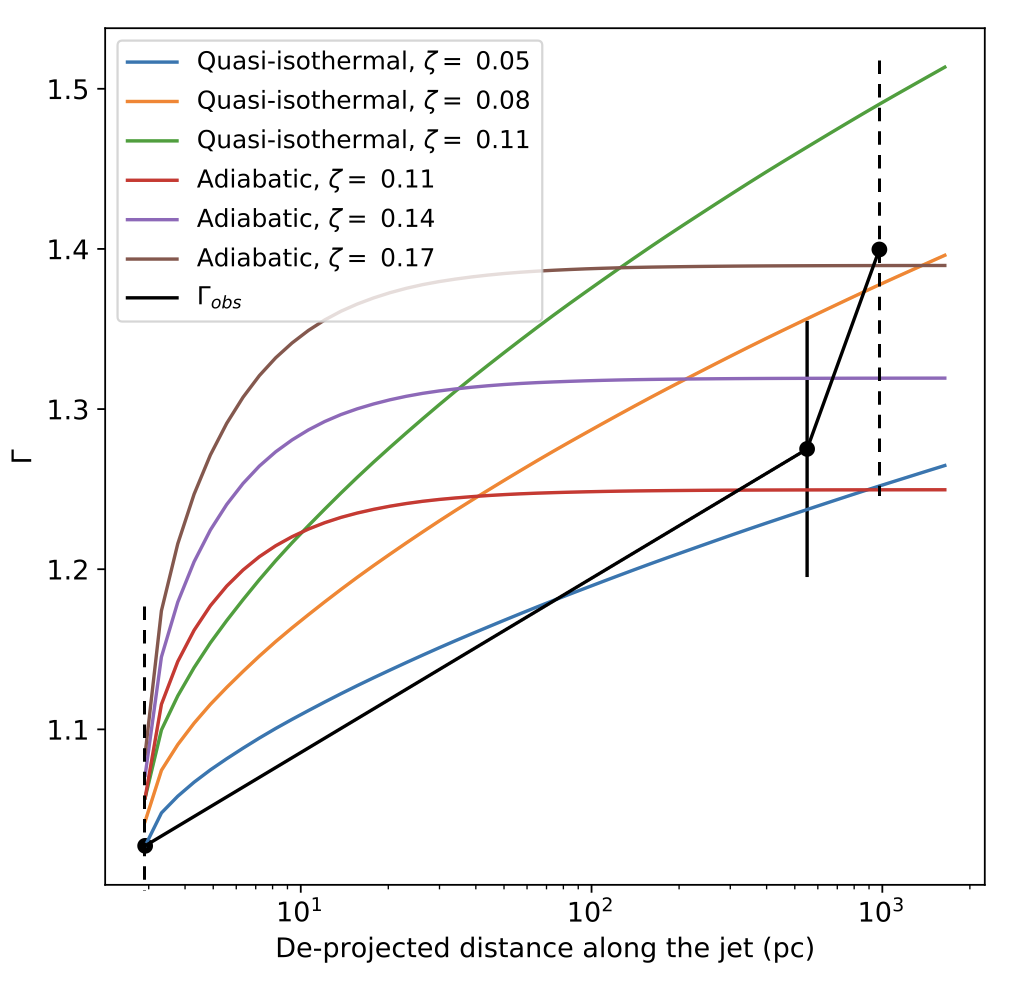}
    \caption{The jet Lorentz factor profile, observed and modelled using different thermal acceleration mechanisms. The observed Lorentz factors of the VLBI and VLA knots have been given in solid points, joined by a black solid line. The error bars for the VLBI and the farthest knots are highly uncertain and therefore have been given in approximate dotted lines. The adiabatic acceleration profiles are given in solid red, magenta and brown lines corresponding to $\zeta=0.11,\,0.14,\,0.17$. The red, orange and green lines correspond to that for the quasi-isothermal case for $\zeta=0.05,\,0.08,\,0.11$. }
    \label{fig:thermal_acc}
\end{figure}

We solved the above equations for an initially non-relativistic fluid with $z_0=1/\sin(20^\circ)$ pc (corresponding to the VLBI knot position),  $\zeta\sim0.05-0.17$ for the adiabatic and quasi-isothermal cases at $\gamma=5/3$, to obtain $\Gamma(z)$, given in solid lines of different colour. Low values of $\zeta$ are plausible since the rest mass energy is mainly dominated by cold non-relativistic protons. The Lorentz factor profiles have been illustrated in Figure \ref{fig:thermal_acc}, alongside the predicted Lorentz factors of the VLBI and VLA knots, assuming the pattern speed \textit{directly} represents the bulk velocity $\beta$. While the adiabatic profile saturates at $\lesssim20$ parsecs, the quasi-isothermal profile continues increasing beyond the terminal Lorentz factor, through $\sim$ few hundred parsec-scales. \textit{Only} using the \textit{observed} Lorentz factors and neglecting the possibility of deceleration in the meso-scales, our results favour a quasi-isothermal model for the bulk acceleration of the 3C 78 jet. However, we cannot prefer either model definitively due to the prevailing uncertainty in our determination of the true bulk speeds.

The other final possibility is that of a magnetic bulk acceleration in the meso-scales, where we have little idea about the jet profile, followed by a deceleration through usual mixing with the interstellar medium or a wind \citep[e.g.,][]{duffell18}, to observed speeds at the few hundred parsec scales. Either a conical profile is maintained throughout, or the jet undergoes a transition from conical to parabolic to conical, depending on the jet environment and its total internal energy. 



\section{Conclusions}
\label{sec:conc}
In this paper we have introduced CAgNVAS, a long-term project that would utilize the richness of the VLA archives and the serendipitous large time baseline ($\gtrsim40$ years) to provide the best possible estimates of large-scale proper motions in jets from active galactic nuclei. Proper motion measurements are the least model-dependent method to obtain an estimate of the energy content of the jet, which is thereafter highly relevant in the context of jet feedback and galaxy evolution studies. The source we studied here was 3C 78, and we obtained the following results after using four epochs of resolution-matched data from the VLA archive:

1. We found that the jet of 3C 78 contains two bright compact Knots B and C. Knot B undergoes mildly relativistic outward motion $\sim0.51c$, while Knot C moves \textit{\rm inward} superluminally at $\sim2.5c$ before moving outward at $\sim0.5c$ between 2000 and 2019. The transverse motion for both the knots was consistent with zero.

2. We have modelled Knot C as the site of a stationary shock, containing a magnetic field whose pitch angle varies with distance over $\sim50-100$ pc, consistent with the space between Knot B and C. The time evolution of the resulting peak in synchrotron emissivity in the post-shocked plasma closely follows the observed motion of Knot C, given mostly plausible model parameters. This allowed us to estimate a possible post-shocked fluid speed $\sim0.7c$ for Knot C. This method can be generalized to the case of moving shocks at various wavelengths.

3. We found that the apparent speeds of Knots B and C, are $\sim5-7$ times larger than the maximum apparent VLBI speed $0.1c$. Using basic estimates from Lorentz velocity addition and fluid dynamics, we deduced that the jet was either indeed undergoing bulk acceleration at the hundred parsec scales or decelerating from an accelerating phase that occurred in the meso-scale between VLBI and VLA.

4. The jet profile remains approximately conical from parsec to kilo-parsec scales.

5. The spectral energy distribution of this source showed two distinct synchrotron components, reminiscent of a "multi-spectral component" jet. The different spectral slopes could be explained by a physically motivated two-zone model (or an "effective" one-zone model) containing shocked plasma and its adiabatically expanded version. We found that the soft X-rays could be produced by IC/CMB, while a part of the $\gamma$-rays was definitely synchrotron self-Compton emission, both of which can be further tested with future monitoring and more sophisticated modelling. We also found that the jet is highly matter-dominated at the hundred-parsec scales.

6. From our observations, we find it unlikely that magnetic acceleration models for the jet acceleration \textit{beyond parsec scales} are viable since there is no observed transition in jet shape and the jet is matter-dominated at the hundred-parsec scales \citep{kov20}. We find that a range of "quasi-isothermal" and adiabatic models of an expanding conical jet from \cite{crumley17} can very crudely produce the observed magnitudes of the jet velocity profile. This would require further testing with deeper observations of this source at better resolution, like with the HST or James Webb Space Telescope (JWST).

\section*{Acknowledgements}

ARC thanks Alan Marscher for insightful comments on proper motions. ARC and ETM thank the anonymous referee for their comments which helped improve the paper. ARC and ETM acknowledge National Science Foundation (NSF) Grant 12971 that supported this work. This paper makes use of the following ALMA data: ADS/JAO.ALMA\#2018.1.00585. ALMA is a partnership of ESO (representing its member states), NSF (USA) and NINS (Japan), together with NRC (Canada), MOST and ASIAA (Taiwan), and KASI (Republic of Korea), in cooperation with the Republic of Chile. The Joint ALMA Observatory is operated by ESO, AUI/NRAO and NAOJ. The National Radio Astronomy Observatory is a facility of the National Science Foundation operated under cooperative agreement by Associated Universities, Inc.

\section*{Data Availability}

The inclusion of a Data Availability Statement is a requirement for articles published in MNRAS. Data Availability Statements provide a standardised format for readers to understand the availability of data underlying the research results described in the article. The statement may refer to original data generated in the course of the study or to third-party data analysed in the article. The statement should describe and provide means of access, where possible, by linking to the data or providing the required accession numbers for the relevant databases or DOIs.



\bibliographystyle{mnras}
\bibliography{version1,from_prop}{} 

\begin{thebibliography}{}
\makeatletter
\relax
\def\mn@urlcharsother{\let\do\@makeother \do\$\do\&\do\#\do\^\do\_\do\%\do\~}
\def\mn@doi{\begingroup\mn@urlcharsother \@ifnextchar [ {\mn@doi@}
  {\mn@doi@[]}}
\def\mn@doi@[#1]#2{\def\@tempa{#1}\ifx\@tempa\@empty \href
  {http://dx.doi.org/#2} {doi:#2}\else \href {http://dx.doi.org/#2} {#1}\fi
  \endgroup}
\def\mn@eprint#1#2{\mn@eprint@#1:#2::\@nil}
\def\mn@eprint@arXiv#1{\href {http://arxiv.org/abs/#1} {{\tt arXiv:#1}}}
\def\mn@eprint@dblp#1{\href {http://dblp.uni-trier.de/rec/bibtex/#1.xml}
  {dblp:#1}}
\def\mn@eprint@#1:#2:#3:#4\@nil{\def\@tempa {#1}\def\@tempb {#2}\def\@tempc
  {#3}\ifx \@tempc \@empty \let \@tempc \@tempb \let \@tempb \@tempa \fi \ifx
  \@tempb \@empty \def\@tempb {arXiv}\fi \@ifundefined
  {mn@eprint@\@tempb}{\@tempb:\@tempc}{\expandafter \expandafter \csname
  mn@eprint@\@tempb\endcsname \expandafter{\@tempc}}}

\bibitem[\protect\citeauthoryear{{Allen} et~al.,}{{Allen}
  et~al.}{2002}]{allen02}
{Allen} M.~G.,  et~al., 2002, \mn@doi [\apjs] {10.1086/338823}, \href
  {https://ui.adsabs.harvard.edu/abs/2002ApJS..139..411A} {139, 411}

\bibitem[\protect\citeauthoryear{{Asada}, {Nakamura}, {Doi}, {Nagai}  \&
  {Inoue}}{{Asada} et~al.}{2014}]{asada14}
{Asada} K.,  {Nakamura} M.,  {Doi} A.,  {Nagai} H.,   {Inoue} M.,  2014,
  \mn@doi [\apjl] {10.1088/2041-8205/781/1/L2}, \href
  {https://ui.adsabs.harvard.edu/abs/2014ApJ...781L...2A} {781, L2}

\bibitem[\protect\citeauthoryear{{Balmaverde} et~al.,}{{Balmaverde}
  et~al.}{2012}]{balma12}
{Balmaverde} B.,  et~al., 2012, \mn@doi [\aap] {10.1051/0004-6361/201219561},
  \href {https://ui.adsabs.harvard.edu/abs/2012A&A...545A.143B} {545, A143}

\bibitem[\protect\citeauthoryear{{Baum} et~al.,}{{Baum} et~al.}{1997}]{baum97}
{Baum} S.~A.,  et~al., 1997, \mn@doi [\apj] {10.1086/304221}, \href
  {https://ui.adsabs.harvard.edu/abs/1997ApJ...483..178B} {483, 178}

\bibitem[\protect\citeauthoryear{{Begelman}, {Blandford}  \& {Rees}}{{Begelman}
  et~al.}{1984}]{begelman84}
{Begelman} M.~C.,  {Blandford} R.~D.,   {Rees} M.~J.,  1984, \mn@doi [Reviews
  of Modern Physics] {10.1103/RevModPhys.56.255}, \href
  {https://ui.adsabs.harvard.edu/abs/1984RvMP...56..255B} {56, 255}

\bibitem[\protect\citeauthoryear{{Bicknell} \& {Begelman}}{{Bicknell} \&
  {Begelman}}{1996}]{bb96}
{Bicknell} G.~V.,  {Begelman} M.~C.,  1996, \mn@doi [\apj] {10.1086/177636},
  \href {https://ui.adsabs.harvard.edu/abs/1996ApJ...467..597B} {467, 597}

\bibitem[\protect\citeauthoryear{{Biretta}, {Zhou}  \& {Owen}}{{Biretta}
  et~al.}{1995}]{biret95_hst}
{Biretta} J.~A.,  {Zhou} F.,   {Owen} F.~N.,  1995, \mn@doi [\apj]
  {10.1086/175901}, \href
  {https://ui.adsabs.harvard.edu/abs/1995ApJ...447..582B} {447, 582}

\bibitem[\protect\citeauthoryear{{Biretta}, {Sparks}  \& {Macchetto}}{{Biretta}
  et~al.}{1999}]{biret99}
{Biretta} J.~A.,  {Sparks} W.~B.,   {Macchetto} F.,  1999, \mn@doi [\apj]
  {10.1086/307499}, \href
  {https://ui.adsabs.harvard.edu/abs/1999ApJ...520..621B} {520, 621}

\bibitem[\protect\citeauthoryear{{Blandford}}{{Blandford}}{1990}]{blandford90}
{Blandford} R.~D.,  1990, in {Blandford} R.~D.,  {Netzer} H.,  {Woltjer} L.,
  {Courvoisier} T.~J.~L.,   {Mayor} M.,  eds, Active Galactic Nuclei. pp
  161--275

\bibitem[\protect\citeauthoryear{{Blandford} \& {K{\"o}nigl}}{{Blandford} \&
  {K{\"o}nigl}}{1979}]{bland79}
{Blandford} R.~D.,  {K{\"o}nigl} A.,  1979, \mn@doi [\apj] {10.1086/157262},
  \href {https://ui.adsabs.harvard.edu/abs/1979ApJ...232...34B} {232, 34}

\bibitem[\protect\citeauthoryear{{Blandford} \& {McKee}}{{Blandford} \&
  {McKee}}{1976}]{bland76}
{Blandford} R.~D.,  {McKee} C.~F.,  1976, \mn@doi [Physics of Fluids]
  {10.1063/1.861619}, \href
  {https://ui.adsabs.harvard.edu/abs/1976PhFl...19.1130B} {19, 1130}

\bibitem[\protect\citeauthoryear{{Blandford} \& {Payne}}{{Blandford} \&
  {Payne}}{1982}]{bland82}
{Blandford} R.~D.,  {Payne} D.~G.,  1982, \mn@doi [\mnras]
  {10.1093/mnras/199.4.883}, \href
  {https://ui.adsabs.harvard.edu/abs/1982MNRAS.199..883B} {199, 883}

\bibitem[\protect\citeauthoryear{{Blandford} \& {Rees}}{{Blandford} \&
  {Rees}}{1978}]{bland78}
{Blandford} R.~D.,  {Rees} M.~J.,  1978, \mn@doi [\physscr]
  {10.1088/0031-8949/17/3/020}, \href
  {https://ui.adsabs.harvard.edu/abs/1978PhyS...17..265B} {17, 265}

\bibitem[\protect\citeauthoryear{{Blandford} \& {Znajek}}{{Blandford} \&
  {Znajek}}{1977}]{blandford77}
{Blandford} R.~D.,  {Znajek} R.~L.,  1977, \mn@doi [\mnras]
  {10.1093/mnras/179.3.433}, \href
  {https://ui.adsabs.harvard.edu/abs/1977MNRAS.179..433B} {179, 433}

\bibitem[\protect\citeauthoryear{{Blandford}, {Meier}  \&
  {Readhead}}{{Blandford} et~al.}{2019}]{blandford2019}
{Blandford} R.,  {Meier} D.,   {Readhead} A.,  2019, \mn@doi [\araa]
  {10.1146/annurev-astro-081817-051948}, \href
  {https://ui.adsabs.harvard.edu/abs/2019ARA&A..57..467B} {57, 467}

\bibitem[\protect\citeauthoryear{{Briggs}, {Schwab}  \& {Sramek}}{{Briggs}
  et~al.}{1999}]{briggs99}
{Briggs} D.~S.,  {Schwab} F.~R.,   {Sramek} R.~A.,  1999, in {Taylor} G.~B.,
  {Carilli} C.~L.,   {Perley} R.~A.,  eds,  Astronomical Society of the Pacific
  Conference Series Vol. 180, Synthesis Imaging in Radio Astronomy II. p.~127

\bibitem[\protect\citeauthoryear{{Broderick} \& {Loeb}}{{Broderick} \&
  {Loeb}}{2009}]{brod09}
{Broderick} A.~E.,  {Loeb} A.,  2009, \mn@doi [\apjl]
  {10.1088/0004-637X/703/2/L104}, \href
  {https://ui.adsabs.harvard.edu/abs/2009ApJ...703L.104B} {703, L104}

\bibitem[\protect\citeauthoryear{{Celotti}, {Ghisellini}  \&
  {Chiaberge}}{{Celotti} et~al.}{2001}]{cel01}
{Celotti} A.,  {Ghisellini} G.,   {Chiaberge} M.,  2001, \mn@doi [\mnras]
  {10.1046/j.1365-8711.2001.04160.x}, \href
  {https://ui.adsabs.harvard.edu/abs/2001MNRAS.321L...1C} {321, L1}

\bibitem[\protect\citeauthoryear{{Cheung}, {Harris}  \& {Stawarz}}{{Cheung}
  et~al.}{2007}]{cheung07}
{Cheung} C.~C.,  {Harris} D.~E.,   {Stawarz} {\L}.,  2007, \mn@doi [\apjl]
  {10.1086/520510}, \href
  {https://ui.adsabs.harvard.edu/abs/2007ApJ...663L..65C} {663, L65}

\bibitem[\protect\citeauthoryear{{Chiaberge}, {Macchetto}, {Sparks}, {Capetti},
  {Allen}  \& {Martel}}{{Chiaberge} et~al.}{2002}]{chia02}
{Chiaberge} M.,  {Macchetto} F.~D.,  {Sparks} W.~B.,  {Capetti} A.,  {Allen}
  M.~G.,   {Martel} A.~R.,  2002, \mn@doi [\apj] {10.1086/339846}, \href
  {https://ui.adsabs.harvard.edu/abs/2002ApJ...571..247C} {571, 247}

\bibitem[\protect\citeauthoryear{{Colina} \& {Perez-Fournon}}{{Colina} \&
  {Perez-Fournon}}{1990}]{colina90}
{Colina} L.,  {Perez-Fournon} I.,  1990, \mn@doi [\apjs] {10.1086/191408},
  \href {https://ui.adsabs.harvard.edu/abs/1990ApJS...72...41C} {72, 41}

\bibitem[\protect\citeauthoryear{{Crumley}, {Ceccobello}, {Connors}  \&
  {Cavecchi}}{{Crumley} et~al.}{2017}]{crumley17}
{Crumley} P.,  {Ceccobello} C.,  {Connors} R. M.~T.,   {Cavecchi} Y.,  2017,
  \mn@doi [\aap] {10.1051/0004-6361/201630229}, \href
  {https://ui.adsabs.harvard.edu/abs/2017A&A...601A..87C} {601, A87}

\bibitem[\protect\citeauthoryear{{Daly} \& {Marscher}}{{Daly} \&
  {Marscher}}{1988}]{daly88}
{Daly} R.~A.,  {Marscher} A.~P.,  1988, \mn@doi [\apj] {10.1086/166858}, \href
  {https://ui.adsabs.harvard.edu/abs/1988ApJ...334..539D} {334, 539}

\bibitem[\protect\citeauthoryear{{Duffell} \& {Laskar}}{{Duffell} \&
  {Laskar}}{2018}]{duffell18}
{Duffell} P.~C.,  {Laskar} T.,  2018, \mn@doi [\apj]
  {10.3847/1538-4357/aadb9c}, \href
  {https://ui.adsabs.harvard.edu/abs/2018ApJ...865...94D} {865, 94}

\bibitem[\protect\citeauthoryear{{Fabian}}{{Fabian}}{2012}]{fabian12}
{Fabian} A.~C.,  2012, \mn@doi [\araa] {10.1146/annurev-astro-081811-125521},
  \href {https://ui.adsabs.harvard.edu/abs/2012ARA&A..50..455F} {50, 455}

\bibitem[\protect\citeauthoryear{{Falcke}, {Malkan}  \& {Biermann}}{{Falcke}
  et~al.}{1995}]{falcke95}
{Falcke} H.,  {Malkan} M.~A.,   {Biermann} P.~L.,  1995, \aap, \href
  {https://ui.adsabs.harvard.edu/abs/1995A&A...298..375F} {298, 375}

\bibitem[\protect\citeauthoryear{{Falle}}{{Falle}}{1991}]{falle91}
{Falle} S.~A.~E.~G.,  1991, \mn@doi [\mnras] {10.1093/mnras/250.3.581}, \href
  {https://ui.adsabs.harvard.edu/abs/1991MNRAS.250..581F} {250, 581}

\bibitem[\protect\citeauthoryear{{Fanaroff} \& {Riley}}{{Fanaroff} \&
  {Riley}}{1974}]{fan74}
{Fanaroff} B.~L.,  {Riley} J.~M.,  1974, \mn@doi [\mnras]
  {10.1093/mnras/167.1.31P}, \href
  {https://ui.adsabs.harvard.edu/abs/1974MNRAS.167P..31F} {167, 31P}

\bibitem[\protect\citeauthoryear{{Fukazawa}, {Finke}, {Stawarz}, {Tanaka},
  {Itoh}  \& {Tokuda}}{{Fukazawa} et~al.}{2015}]{fukazawa15}
{Fukazawa} Y.,  {Finke} J.,  {Stawarz} {\L}.,  {Tanaka} Y.,  {Itoh} R.,
  {Tokuda} S.,  2015, \mn@doi [\apj] {10.1088/0004-637X/798/2/74}, \href
  {https://ui.adsabs.harvard.edu/abs/2015ApJ...798...74F} {798, 74}

\bibitem[\protect\citeauthoryear{{Gaibler}, {Krause}  \& {Camenzind}}{{Gaibler}
  et~al.}{2009}]{gaibler09}
{Gaibler} V.,  {Krause} M.,   {Camenzind} M.,  2009, \mn@doi [\mnras]
  {10.1111/j.1365-2966.2009.15625.x}, \href
  {https://ui.adsabs.harvard.edu/abs/2009MNRAS.400.1785G} {400, 1785}

\bibitem[\protect\citeauthoryear{{Georganopoulos}, {Kazanas}, {Perlman}  \&
  {Stecker}}{{Georganopoulos} et~al.}{2005}]{georg05}
{Georganopoulos} M.,  {Kazanas} D.,  {Perlman} E.,   {Stecker} F.~W.,  2005,
  \mn@doi [\apj] {10.1086/429558}, \href
  {https://ui.adsabs.harvard.edu/abs/2005ApJ...625..656G} {625, 656}

\bibitem[\protect\citeauthoryear{{Hardcastle} \& {Croston}}{{Hardcastle} \&
  {Croston}}{2020}]{hardcast20}
{Hardcastle} M.~J.,  {Croston} J.~H.,  2020, \mn@doi [\nar]
  {10.1016/j.newar.2020.101539}, \href
  {https://ui.adsabs.harvard.edu/abs/2020NewAR..8801539H} {88, 101539}

\bibitem[\protect\citeauthoryear{{Harris} \& {Krawczynski}}{{Harris} \&
  {Krawczynski}}{2006}]{harris06}
{Harris} D.~E.,  {Krawczynski} H.,  2006, \mn@doi [\araa]
  {10.1146/annurev.astro.44.051905.092446}, \href
  {https://ui.adsabs.harvard.edu/abs/2006ARA&A..44..463H} {44, 463}

\bibitem[\protect\citeauthoryear{{Homan}, {Ojha}, {Wardle}, {Roberts}, {Aller},
  {Aller}  \& {Hughes}}{{Homan} et~al.}{2001}]{homan01}
{Homan} D.~C.,  {Ojha} R.,  {Wardle} J. F.~C.,  {Roberts} D.~H.,  {Aller}
  M.~F.,  {Aller} H.~D.,   {Hughes} P.~A.,  2001, \mn@doi [\apj]
  {10.1086/319466}, \href
  {https://ui.adsabs.harvard.edu/abs/2001ApJ...549..840H} {549, 840}

\bibitem[\protect\citeauthoryear{{Jorstad}, {Marscher}, {Mattox}, {Wehrle},
  {Bloom}  \& {Yurchenko}}{{Jorstad} et~al.}{2001}]{jorstad01}
{Jorstad} S.~G.,  {Marscher} A.~P.,  {Mattox} J.~R.,  {Wehrle} A.~E.,  {Bloom}
  S.~D.,   {Yurchenko} A.~V.,  2001, \mn@doi [\apjs] {10.1086/320858}, \href
  {https://ui.adsabs.harvard.edu/abs/2001ApJS..134..181J} {134, 181}

\bibitem[\protect\citeauthoryear{{Junor} \& {Biretta}}{{Junor} \&
  {Biretta}}{1995}]{junor95}
{Junor} W.,  {Biretta} J.~A.,  1995, \mn@doi [\aj] {10.1086/117295}, \href
  {https://ui.adsabs.harvard.edu/abs/1995AJ....109..500J} {109, 500}

\bibitem[\protect\citeauthoryear{{Kataoka} et~al.,}{{Kataoka}
  et~al.}{2008}]{kat08}
{Kataoka} J.,  et~al., 2008, \mn@doi [\apj] {10.1086/523093}, \href
  {https://ui.adsabs.harvard.edu/abs/2008ApJ...672..787K} {672, 787}

\bibitem[\protect\citeauthoryear{{King} \& {Pounds}}{{King} \&
  {Pounds}}{2015}]{king15}
{King} A.,  {Pounds} K.,  2015, \mn@doi [\araa]
  {10.1146/annurev-astro-082214-122316}, \href
  {https://ui.adsabs.harvard.edu/abs/2015ARA&A..53..115K} {53, 115}

\bibitem[\protect\citeauthoryear{{Komissarov}, {Barkov}, {Vlahakis}  \&
  {K{\"o}nigl}}{{Komissarov} et~al.}{2007}]{komis07}
{Komissarov} S.~S.,  {Barkov} M.~V.,  {Vlahakis} N.,   {K{\"o}nigl} A.,  2007,
  \mn@doi [\mnras] {10.1111/j.1365-2966.2007.12050.x}, \href
  {https://ui.adsabs.harvard.edu/abs/2007MNRAS.380...51K} {380, 51}

\bibitem[\protect\citeauthoryear{{Komissarov}, {Vlahakis}, {K{\"o}nigl}  \&
  {Barkov}}{{Komissarov} et~al.}{2009}]{komiss09}
{Komissarov} S.~S.,  {Vlahakis} N.,  {K{\"o}nigl} A.,   {Barkov} M.~V.,  2009,
  \mn@doi [\mnras] {10.1111/j.1365-2966.2009.14410.x}, \href
  {https://ui.adsabs.harvard.edu/abs/2009MNRAS.394.1182K} {394, 1182}

\bibitem[\protect\citeauthoryear{{Konigl}}{{Konigl}}{1980}]{konig80}
{Konigl} A.,  1980, \mn@doi [Physics of Fluids] {10.1063/1.863110}, \href
  {https://ui.adsabs.harvard.edu/abs/1980PhFl...23.1083K} {23, 1083}

\bibitem[\protect\citeauthoryear{{Kovalev}, {Lister}, {Homan}  \&
  {Kellermann}}{{Kovalev} et~al.}{2007}]{kovalev07}
{Kovalev} Y.~Y.,  {Lister} M.~L.,  {Homan} D.~C.,   {Kellermann} K.~I.,  2007,
  \mn@doi [\apjl] {10.1086/522603}, \href
  {https://ui.adsabs.harvard.edu/abs/2007ApJ...668L..27K} {668, L27}

\bibitem[\protect\citeauthoryear{{Kovalev}, {Pushkarev}, {Nokhrina}, {Plavin},
  {Beskin}, {Chernoglazov}, {Lister}  \& {Savolainen}}{{Kovalev}
  et~al.}{2020}]{kov20}
{Kovalev} Y.~Y.,  {Pushkarev} A.~B.,  {Nokhrina} E.~E.,  {Plavin} A.~V.,
  {Beskin} V.~S.,  {Chernoglazov} A.~V.,  {Lister} M.~L.,   {Savolainen} T.,
  2020, \mn@doi [\mnras] {10.1093/mnras/staa1121}, \href
  {https://ui.adsabs.harvard.edu/abs/2020MNRAS.495.3576K} {495, 3576}

\bibitem[\protect\citeauthoryear{{Leahy}}{{Leahy}}{1991}]{leahy91}
{Leahy} J.~P.,  1991, {Interpretation of large scale extragalactic Jets, Beams
  and Jets in Astrophysics. Edited by P.A. Hughes}.
p.~100

\bibitem[\protect\citeauthoryear{{Lind} \& {Blandford}}{{Lind} \&
  {Blandford}}{1985}]{lind85}
{Lind} K.~R.,  {Blandford} R.~D.,  1985, \mn@doi [\apj] {10.1086/163380}, \href
  {https://ui.adsabs.harvard.edu/abs/1985ApJ...295..358L} {295, 358}

\bibitem[\protect\citeauthoryear{{Lister} et~al.,}{{Lister}
  et~al.}{2016}]{lister16}
{Lister} M.~L.,  et~al., 2016, \mn@doi [\aj] {10.3847/0004-6256/152/1/12},
  \href {https://ui.adsabs.harvard.edu/abs/2016AJ....152...12L} {152, 12}

\bibitem[\protect\citeauthoryear{{Lister}, {Aller}, {Aller}, {Hodge}, {Homan},
  {Kovalev}, {Pushkarev}  \& {Savolainen}}{{Lister} et~al.}{2018}]{lister18}
{Lister} M.~L.,  {Aller} M.~F.,  {Aller} H.~D.,  {Hodge} M.~A.,  {Homan} D.~C.,
   {Kovalev} Y.~Y.,  {Pushkarev} A.~B.,   {Savolainen} T.,  2018, \mn@doi
  [\apjs] {10.3847/1538-4365/aa9c44}, \href
  {https://ui.adsabs.harvard.edu/abs/2018ApJS..234...12L} {234, 12}

\bibitem[\protect\citeauthoryear{{Lister} et~al.,}{{Lister}
  et~al.}{2019}]{lister19}
{Lister} M.~L.,  et~al., 2019, \mn@doi [\apj] {10.3847/1538-4357/ab08ee}, \href
  {https://ui.adsabs.harvard.edu/abs/2019ApJ...874...43L} {874, 43}

\bibitem[\protect\citeauthoryear{{Liu} \& {Xie}}{{Liu} \& {Xie}}{1992}]{liu92}
{Liu} F.~K.,  {Xie} G.~Z.,  1992, \aaps, \href
  {https://ui.adsabs.harvard.edu/abs/1992A&AS...95..249L} {95, 249}

\bibitem[\protect\citeauthoryear{{Ly}, {Walker}  \& {Junor}}{{Ly}
  et~al.}{2007}]{ly2007}
{Ly} C.,  {Walker} R.~C.,   {Junor} W.,  2007, \mn@doi [\apj] {10.1086/512846},
  \href {https://ui.adsabs.harvard.edu/abs/2007ApJ...660..200L} {660, 200}

\bibitem[\protect\citeauthoryear{{Martel} et~al.,}{{Martel}
  et~al.}{1999}]{martel99}
{Martel} A.~R.,  et~al., 1999, \mn@doi [\apjs] {10.1086/313205}, \href
  {https://ui.adsabs.harvard.edu/abs/1999ApJS..122...81M} {122, 81}

\bibitem[\protect\citeauthoryear{{Massaro} et~al.,}{{Massaro}
  et~al.}{2015}]{massaro15}
{Massaro} F.,  et~al., 2015, \mn@doi [\apjs] {10.1088/0067-0049/220/1/5}, \href
  {https://ui.adsabs.harvard.edu/abs/2015ApJS..220....5M} {220, 5}

\bibitem[\protect\citeauthoryear{{McKinney} \& {Blandford}}{{McKinney} \&
  {Blandford}}{2009}]{mckinney09}
{McKinney} J.~C.,  {Blandford} R.~D.,  2009, \mn@doi [\mnras]
  {10.1111/j.1745-3933.2009.00625.x}, \href
  {https://ui.adsabs.harvard.edu/abs/2009MNRAS.394L.126M} {394, L126}

\bibitem[\protect\citeauthoryear{{McMullin}, {Waters}, {Schiebel}, {Young}  \&
  {Golap}}{{McMullin} et~al.}{2007}]{casa}
{McMullin} J.~P.,  {Waters} B.,  {Schiebel} D.,  {Young} W.,   {Golap} K.,
  2007, in {Shaw} R.~A.,  {Hill} F.,   {Bell} D.~J.,  eds,  Astronomical
  Society of the Pacific Conference Series Vol. 376, Astronomical Data Analysis
  Software and Systems XVI. p.~127

\bibitem[\protect\citeauthoryear{{Mehta}, {Georganopoulos}, {Perlman},
  {Padgett}  \& {Chartas}}{{Mehta} et~al.}{2009}]{mehta09}
{Mehta} K.~T.,  {Georganopoulos} M.,  {Perlman} E.~S.,  {Padgett} C.~A.,
  {Chartas} G.,  2009, \mn@doi [\apj] {10.1088/0004-637X/690/2/1706}, \href
  {https://ui.adsabs.harvard.edu/abs/2009ApJ...690.1706M} {690, 1706}

\bibitem[\protect\citeauthoryear{{Meyer}, {Sparks}, {Biretta}, {Anderson},
  {Sohn}, {van der Marel}, {Norman}  \& {Nakamura}}{{Meyer}
  et~al.}{2013}]{meyer13}
{Meyer} E.~T.,  {Sparks} W.~B.,  {Biretta} J.~A.,  {Anderson} J.,  {Sohn}
  S.~T.,  {van der Marel} R.~P.,  {Norman} C.,   {Nakamura} M.,  2013, \mn@doi
  [\apjl] {10.1088/2041-8205/774/2/L21}, \href
  {https://ui.adsabs.harvard.edu/abs/2013ApJ...774L..21M} {774, L21}

\bibitem[\protect\citeauthoryear{{Meyer}, {Georganopoulos}, {Sparks},
  {Godfrey}, {Lovell}  \& {Perlman}}{{Meyer} et~al.}{2015}]{meyer15}
{Meyer} E.~T.,  {Georganopoulos} M.,  {Sparks} W.~B.,  {Godfrey} L.,  {Lovell}
  J. E.~J.,   {Perlman} E.,  2015, \mn@doi [\apj]
  {10.1088/0004-637X/805/2/154}, \href
  {https://ui.adsabs.harvard.edu/abs/2015ApJ...805..154M} {805, 154}

\bibitem[\protect\citeauthoryear{{Meyer} et~al.,}{{Meyer}
  et~al.}{2016}]{meyer16_3c273}
{Meyer} E.~T.,  et~al., 2016, \mn@doi [\apj] {10.3847/0004-637X/818/2/195},
  \href {https://ui.adsabs.harvard.edu/abs/2016ApJ...818..195M} {818, 195}

\bibitem[\protect\citeauthoryear{{Meyer}, {Petropoulou}, {Georganopoulos},
  {Chiaberge}, {Breiding}  \& {Sparks}}{{Meyer} et~al.}{2018}]{meyer18_m84}
{Meyer} E.~T.,  {Petropoulou} M.,  {Georganopoulos} M.,  {Chiaberge} M.,
  {Breiding} P.,   {Sparks} W.~B.,  2018, \mn@doi [\apj]
  {10.3847/1538-4357/aabf39}, \href
  {https://ui.adsabs.harvard.edu/abs/2018ApJ...860....9M} {860, 9}

\bibitem[\protect\citeauthoryear{{Middelberg} et~al.,}{{Middelberg}
  et~al.}{2004}]{middel04}
{Middelberg} E.,  et~al., 2004, \mn@doi [\aap] {10.1051/0004-6361:20040019},
  \href {https://ui.adsabs.harvard.edu/abs/2004A&A...417..925M} {417, 925}

\bibitem[\protect\citeauthoryear{{M{\"u}ller} et~al.,}{{M{\"u}ller}
  et~al.}{2014}]{muller14_tanami}
{M{\"u}ller} C.,  et~al., 2014, \mn@doi [\aap] {10.1051/0004-6361/201423948},
  \href {https://ui.adsabs.harvard.edu/abs/2014A&A...569A.115M} {569, A115}

\bibitem[\protect\citeauthoryear{{Ohira}, {Yamazaki}, {Kawanaka}  \&
  {Ioka}}{{Ohira} et~al.}{2012}]{ohira12}
{Ohira} Y.,  {Yamazaki} R.,  {Kawanaka} N.,   {Ioka} K.,  2012, \mn@doi
  [\mnras] {10.1111/j.1365-2966.2012.21908.x}, \href
  {https://ui.adsabs.harvard.edu/abs/2012MNRAS.427...91O} {427, 91}

\bibitem[\protect\citeauthoryear{{Pacholczyk}}{{Pacholczyk}}{1970}]{pachol70}
{Pacholczyk} A.~G.,  1970, {Radio astrophysics. Nonthermal processes in
  galactic and extragalactic sources, San Francisco: Freeman}

\bibitem[\protect\citeauthoryear{{Phinney}}{{Phinney}}{1985}]{phinney85}
{Phinney} E.~S.,  1985, in {Miller} J.~S.,  ed., Astrophysics of Active
  Galaxies and Quasi-Stellar Objects. pp 453--496, \url
  {https://articles.adsabs.harvard.edu/pdf/1985aagq.conf..453P}

\bibitem[\protect\citeauthoryear{{Piner} \& {Edwards}}{{Piner} \&
  {Edwards}}{2018}]{piner18}
{Piner} B.~G.,  {Edwards} P.~G.,  2018, \mn@doi [\apj]
  {10.3847/1538-4357/aaa425}, \href
  {https://ui.adsabs.harvard.edu/abs/2018ApJ...853...68P} {853, 68}

\bibitem[\protect\citeauthoryear{{Pushkarev}, {Kovalev}, {Lister}  \&
  {Savolainen}}{{Pushkarev} et~al.}{2017}]{pushkarev17}
{Pushkarev} A.~B.,  {Kovalev} Y.~Y.,  {Lister} M.~L.,   {Savolainen} T.,  2017,
  \mn@doi [\mnras] {10.1093/mnras/stx854}, \href
  {https://ui.adsabs.harvard.edu/abs/2017MNRAS.468.4992P} {468, 4992}

\bibitem[\protect\citeauthoryear{{Reid}, {Biretta}, {Junor}, {Muxlow}  \&
  {Spencer}}{{Reid} et~al.}{1989}]{reid89}
{Reid} M.~J.,  {Biretta} J.~A.,  {Junor} W.,  {Muxlow} T.~W.~B.,   {Spencer}
  R.~E.,  1989, \mn@doi [\apj] {10.1086/166998}, \href
  {https://ui.adsabs.harvard.edu/abs/1989ApJ...336..112R} {336, 112}

\bibitem[\protect\citeauthoryear{{Rivas}, {Arsham}  \&
  {Georganopoulos}}{{Rivas} et~al.}{2014}]{rivas14}
{Rivas} D.,  {Arsham} A.,   {Georganopoulos} M.,  2014, in American
  Astronomical Society Meeting Abstracts \#223. p. 251.14

\bibitem[\protect\citeauthoryear{{Roychowdhury} \& {Meyer}}{{Roychowdhury} \&
  {Meyer}}{2023a}]{royc_evn23}
{Roychowdhury} A.,  {Meyer} E.,  2023a, in 15th European VLBI Network
  Mini-Symposium and Users' Meeting. p.~59

\bibitem[\protect\citeauthoryear{{Roychowdhury} \& {Meyer}}{{Roychowdhury} \&
  {Meyer}}{2023b}]{royc23_arxiv_ngdif}
{Roychowdhury} A.,  {Meyer} E.~T.,  2023b, \mn@doi [arXiv e-prints]
  {10.48550/arXiv.2308.00832}, \href
  {https://ui.adsabs.harvard.edu/abs/2023arXiv230800832R} {p. arXiv:2308.00832}

\bibitem[\protect\citeauthoryear{{Saikia}, {Subrahmanya}, {Patnaik}, {Unger},
  {Cornwell}, {Graham}  \& {Prabhu}}{{Saikia} et~al.}{1986}]{saik86}
{Saikia} D.~J.,  {Subrahmanya} C.~R.,  {Patnaik} A.~R.,  {Unger} S.~W.,
  {Cornwell} T.~J.,  {Graham} D.~A.,   {Prabhu} T.~P.,  1986, \mn@doi [\mnras]
  {10.1093/mnras/219.3.545}, \href
  {https://ui.adsabs.harvard.edu/abs/1986MNRAS.219..545S} {219, 545}

\bibitem[\protect\citeauthoryear{{Schwartz} et~al.,}{{Schwartz}
  et~al.}{2000}]{sch00}
{Schwartz} D.~A.,  et~al., 2000, \mn@doi [\apjl] {10.1086/312875}, \href
  {https://ui.adsabs.harvard.edu/abs/2000ApJ...540L..69S} {540, 69}

\bibitem[\protect\citeauthoryear{{Shepherd}, {Pearson}  \& {Taylor}}{{Shepherd}
  et~al.}{1994}]{shep94}
{Shepherd} M.~C.,  {Pearson} T.~J.,   {Taylor} G.~B.,  1994, in Bulletin of the
  American Astronomical Society. pp 987--989

\bibitem[\protect\citeauthoryear{{Sikora}, {Sol}, {Begelman}  \&
  {Madejski}}{{Sikora} et~al.}{1996}]{sikora96}
{Sikora} M.,  {Sol} H.,  {Begelman} M.~C.,   {Madejski} G.,  1996, \aaps, \href
  {https://ui.adsabs.harvard.edu/abs/1996A&AS..120C.579S} {120, 579}

\bibitem[\protect\citeauthoryear{{Snios} et~al.,}{{Snios}
  et~al.}{2019a}]{snios_cenA}
{Snios} B.,  et~al., 2019a, \mn@doi [\apj] {10.3847/1538-4357/aafaf3}, \href
  {https://ui.adsabs.harvard.edu/abs/2019ApJ...871..248S} {871, 248}

\bibitem[\protect\citeauthoryear{{Snios}, {Nulsen}, {Kraft}, {Cheung}, {Meyer},
  {Forman}, {Jones}  \& {Murray}}{{Snios} et~al.}{2019b}]{snios19}
{Snios} B.,  {Nulsen} P. E.~J.,  {Kraft} R.~P.,  {Cheung} C.~C.,  {Meyer}
  E.~T.,  {Forman} W.~R.,  {Jones} C.,   {Murray} S.~S.,  2019b, \mn@doi [\apj]
  {10.3847/1538-4357/ab2119}, \href
  {https://ui.adsabs.harvard.edu/abs/2019ApJ...879....8S} {879, 8}

\bibitem[\protect\citeauthoryear{{Sparks}, {Golombek}, {Baum}, {Biretta}, {de
  Koff}, {Macchetto}, {McCarthy}  \& {Miley}}{{Sparks} et~al.}{1995}]{sparks95}
{Sparks} W.~B.,  {Golombek} D.,  {Baum} S.~A.,  {Biretta} J.,  {de Koff} S.,
  {Macchetto} F.,  {McCarthy} P.,   {Miley} G.~K.,  1995, \mn@doi [\apjl]
  {10.1086/316777}, \href
  {https://ui.adsabs.harvard.edu/abs/1995ApJ...450L..55S} {450, L55}

\bibitem[\protect\citeauthoryear{{Tavecchio}, {Maraschi}, {Sambruna}  \&
  {Urry}}{{Tavecchio} et~al.}{2000}]{tav00}
{Tavecchio} F.,  {Maraschi} L.,  {Sambruna} R.~M.,   {Urry} C.~M.,  2000,
  \mn@doi [\apjl] {10.1086/317292}, \href
  {https://ui.adsabs.harvard.edu/abs/2000ApJ...544L..23T} {544, L23}

\bibitem[\protect\citeauthoryear{{Tchekhovskoy}, {Narayan}  \&
  {McKinney}}{{Tchekhovskoy} et~al.}{2011}]{tch11}
{Tchekhovskoy} A.,  {Narayan} R.,   {McKinney} J.~C.,  2011, \mn@doi [\mnras]
  {10.1111/j.1745-3933.2011.01147.x}, \href
  {https://ui.adsabs.harvard.edu/abs/2011MNRAS.418L..79T} {418, L79}

\bibitem[\protect\citeauthoryear{{Tingay} et~al.,}{{Tingay}
  et~al.}{1998}]{tingay98}
{Tingay} S.~J.,  et~al., 1998, \mn@doi [\aj] {10.1086/300257}, \href
  {https://ui.adsabs.harvard.edu/abs/1998AJ....115..960T} {115, 960}

\bibitem[\protect\citeauthoryear{{Tremblay} et~al.,}{{Tremblay}
  et~al.}{2009}]{temb09}
{Tremblay} G.~R.,  et~al., 2009, \mn@doi [\apjs] {10.1088/0067-0049/183/2/278},
  \href {https://ui.adsabs.harvard.edu/abs/2009ApJS..183..278T} {183, 278}

\bibitem[\protect\citeauthoryear{{Unger}, {Booler}  \& {Pedlar}}{{Unger}
  et~al.}{1984}]{unger84}
{Unger} S.~W.,  {Booler} R.~V.,   {Pedlar} A.,  1984, \mn@doi [\mnras]
  {10.1093/mnras/207.4.679}, \href
  {https://ui.adsabs.harvard.edu/abs/1984MNRAS.207..679U} {207, 679}

\bibitem[\protect\citeauthoryear{{Urry} \& {Padovani}}{{Urry} \&
  {Padovani}}{1995}]{urry95}
{Urry} C.~M.,  {Padovani} P.,  1995, \mn@doi [\pasp] {10.1086/133630}, \href
  {https://ui.adsabs.harvard.edu/abs/1995PASP..107..803U} {107, 803}

\bibitem[\protect\citeauthoryear{{Walker}}{{Walker}}{1997}]{walk97}
{Walker} R.~C.,  1997, \mn@doi [\apj] {10.1086/304737}, \href
  {https://ui.adsabs.harvard.edu/abs/1997ApJ...488..675W} {488, 675}

\bibitem[\protect\citeauthoryear{{Walker}, {Hardee}, {Davies}, {Ly}  \&
  {Junor}}{{Walker} et~al.}{2018}]{walker18}
{Walker} R.~C.,  {Hardee} P.~E.,  {Davies} F.~B.,  {Ly} C.,   {Junor} W.,
  2018, \mn@doi [\apj] {10.3847/1538-4357/aaafcc}, \href
  {https://ui.adsabs.harvard.edu/abs/2018ApJ...855..128W} {855, 128}

\bibitem[\protect\citeauthoryear{{Zensus}, {Cohen}  \& {Unwin}}{{Zensus}
  et~al.}{1995}]{zensus95}
{Zensus} J.~A.,  {Cohen} M.~H.,   {Unwin} S.~C.,  1995, \mn@doi [\apj]
  {10.1086/175501}, \href
  {https://ui.adsabs.harvard.edu/abs/1995ApJ...443...35Z} {443, 35}

\bibitem[\protect\citeauthoryear{{Zirbel} \& {Baum}}{{Zirbel} \&
  {Baum}}{1998}]{zirbel98}
{Zirbel} E.~L.,  {Baum} S.~A.,  1998, \mn@doi [\apjs] {10.1086/313070}, \href
  {https://ui.adsabs.harvard.edu/abs/1998ApJS..114..177Z} {114, 177}

\makeatother
\end{thebibliography}




\appendix

\section{Modelling of Backward Motion around a Stationary Shock}

Consider Knot C to be the site of a stationary shock, where $\beta'_{\rm shock}$ is constant and $\bm{\beta'_{\rm shock}}\simeq-\bm{\beta_{\rm pre}}$ through $\sim35$ years. Figure \ref{fig:ss} shows the starting model, consisting of a conical jet section (valid assumption, see Section \ref{sec:profile}), a stationary unmagnetized shock, and a non-emitting pre-shocked and an emitting post-shocked fluid, at $t\neq0$, in the lab frame. We have neglected the detailed nature of the shock and the particle acceleration mechanism for the goal of this work. The pre-shocked fluid continually moves through the shock and $z_s$ denotes the position of the unwavering stationary shock and $z_f(t)=z_s+\beta_{\rm post}ct$ denotes the current position of the end of the post-shock fluid. We assume the post-shock fluid, after passing through the shock, undergoes adiabatic expansion and cools (see \citealt{lind85} for a description of other possibilities). At any time $t$ and position $z>z_s$, the synchrotron emissivity $j_\nu(z,t)$ ($\nu_{\rm min}\ll\nu\ll\nu_{\rm max}$) of the post-shock fluid can be determined as follows (using an effectively toroidal non-turbulent magnetic field) \citep[e.g.,][]{pachol70,leahy91,baum97}:

\begin{figure}
    \centering
    \includegraphics[width=\linewidth]{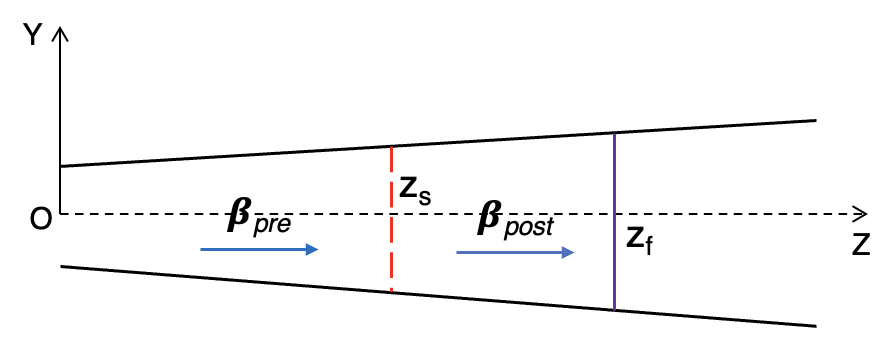}
    \caption{Basic model of flow through a stationary shock inside a conical funnel, with formation of shocked plasma given an endless supply of pre-shocked plasma, as observed in the lab frame at $\theta=\pi/2$. The shock position is marked as $z_s$, with a red dashed line. While the shock may be generally conical and oblique, a plane-perpendicular shock has been used equivalently for a simple analysis. The final position of the post-shocked plasma is marked with a violet line at $z=z_f$.}
    \label{fig:ss}
\end{figure}

\begin{equation}
\begin{split}
\label{eq:jnu}
j_\nu(z,t)\propto K(z_s)\bigg[B(z_s)\sin\Phi(z,t)\bigg]^{\alpha+1}\\
\nu^{-\alpha}\bigg[\frac{r}{r_s}\bigg]^{-3-7\alpha/3}\bigg[\frac{z}{z_s}\bigg]^{-2-5\alpha/3}\delta_{\rm post}^{2+\alpha}(z)\Theta(z_f(t)-z)
\end{split}
\end{equation}

where $K(z_s)$ is the electronic luminosity at $z_s$ in a power-law electron injection $Q\propto K\gamma^{-s}$, $\alpha=(s-1)/2$. $B(z_s)$ is the magnetic field at $z_s$, and $\Phi$ is the angle between the magnetic field and our line of sight, \textit{both measured in the frame of the jet}. $\delta_{\rm post}$ is the Doppler factor of the post-shocked flow. Addition of synchrotron cooling would only make the decrease in $j_{\nu}$ faster; however, for low electronic energies and magnetic fields, cooling due to adiabatic expansion may dominate, especially for $\nu_{\rm min}\ll\nu\ll\nu_{\rm max}$. Note that $j_{\nu}$ is an implicit function of the viewing angle $\theta$ (in the lab frame), which we have discussed in this section. 

Therefore, from Equation \ref{eq:jnu}, it is clear that if $B\sin\Phi$ is constant, the emissivity will peak at $z_s$ for all time. In contrast, if $B\sin\Phi$ is continuously varying, the consequences become more complex. Specifically, if $B\sin\Phi$ completes one period over a feasible length of the jet, a backward motion of the peak emissivity will be possible due to a switch between a minima and maxima of $B\sin\Phi$ at $z_s$. A straightforward phenomenological consideration can be obtained by using $B=$ constant and a linearly increasing $\Phi$ from 0 to 2$\pi$ between $z=0$ and $z=L$, for example. However, we intend to discuss the same more generally. In the co-moving frame of the jet, a general magnetic field with strength $B(z)$ (following $\nabla\cdot\bm{B}=0$ and ignoring decay due to flux freezing since it is included in Equation \ref{eq:jnu}) can be written as a sum of poloidal ($\bm{\hat{z}}$) and toroidal ($\bm{\hat{\phi}}$) components  $\bm{B(z)}=B_p\bm{\hat{z}}+(B_{\phi}\sin\alpha_p/r(z))\bm{\hat{\phi}}$ in cylindrical coordinates $(z,r,\phi)$, where $\alpha_p$ is the magnetic field pitch angle with respect to the jet velocity $\bm{\beta}=\beta\bm{\hat{z}}$. We consider $B_p\ll B_\phi$ since it is expected that the field is dominantly toroidal at kpc-scale from flux freezing. In the same cylindrical coordinates, the electromagnetic wave vector in our direction would be given (see \citealt{brod09} for a detailed description of rotation measures in jets due to different magnetic field configurations) as $\bm{\hat{k}}=\cos\theta'\bm{\hat{z}}+\sin\theta'\cos\phi\bm{\hat{r}}-\sin\theta'\sin\phi\bm{\hat{\phi}}/r$, where $\cos\theta'=(\cos\theta-\beta_{\rm post})/(1-\beta_{\rm post}\cos\theta)$. If $\alpha_p$ increases from $-\pi/2$ to $\pi/2$ along $z$, $\alpha_p(z_f\geq z\geq z_s,t)=\pi/2[1-\{(z_f-z)/z_s\}^m]$. The parameter $m$ encodes the degree of symmetry of the magnetic field lines between $z_s$ and $z_f$ when $z_f=2z_s$, where $m=1$ is the case of $B(z=z_s/2)=B_p$ and $B(z=z_s,2z_s)=B_\phi\sin\alpha_p$. All of the above implies:

\begin{equation}
\begin{split}
\cos\Phi&=\frac{\bm{B}\cdot\bm{\hat{k}}}{|\bm{B}|}=\bigg[\frac{B_p\cos\theta'-B_\phi\sin\alpha_p\sin\theta'\sin\phi}{\sqrt{B_p^2+B_\phi^2\sin^2\alpha_p}}\bigg]\\
\implies & B(z)\sin\Phi\simeq B_\phi\sin\alpha_p\sin[\cos^{-1}(-\sin\theta'\sin\phi)]\\&=B_\phi\sin\alpha_p\sqrt{1-\sin^2\theta'\sin^2\phi}
\end{split}
\end{equation}

where the second step results from $B_\phi\gg B_p$ and the fact that $B_p\sim0$ independent of $\alpha_p$. At $z=z_s$ and $t=0$, $\alpha_p=0$, or $B(z_s)\sin\Phi=0$. However, $\phi\in[0,2\pi]$ and therefore we replace $B(z_s)\sin\Phi$ with $B_\phi\sin\alpha_p\langle\sin\Phi\rangle$, where the average is done over $\phi$. Such a configuration will have a distinct polarization pattern across $z$ and must be testable. We are planning to discuss this in a separate paper as it is out of the scope of this work. In addition to the magnetic field and its orientation, Equation \ref{eq:jnu} must also incorporate $j_{\nu}$ as a function of projected distance $z$ and observed time $t$. 

The basic transformation equations between one-dimensional motion in the jet frame and the lab frame can be written as follows for viewing angle $\theta$:

\begin{equation}
\begin{split}
dz&=dz'\sin\theta,\\
dt&=dt'\big[1-\frac{dz'(t')}{cdt'}\cos\theta\big]\\
\implies t&=t'-\frac{[z'(t')-z'(0)]}{c}\cos\theta,\\
\frac{dz}{dt}(t')&=\frac{dz'}{dt'}\frac{dt'}{dt}\sin\theta=\frac{\dot{z}'(t')\sin\theta}{1-\dot{z}'(t')\cos\theta},\\
\end{split}
\end{equation}
where the primed variables are jet rest-frame quantities and $t=f(t')$, $z(t)=z(f(t'))$ and 
$\dot{z}(t)=\dot{z}(f(t'))$. Using the above transformations, and convolving $j_\nu(z,t)$ with a beam size of desired resolution, it is directly possible to obtain the evolution of the peak emissivity position with time and its velocity, as a function of $\theta$, $\beta_{\rm post}$ or $\alpha$. For our purpose, we use a beam size of width $0.1"$, corresponding to the VLA K-Band A-config (22 GHz) resolution. 

\begin{figure*}
    \centering
    \includegraphics[width=\linewidth]{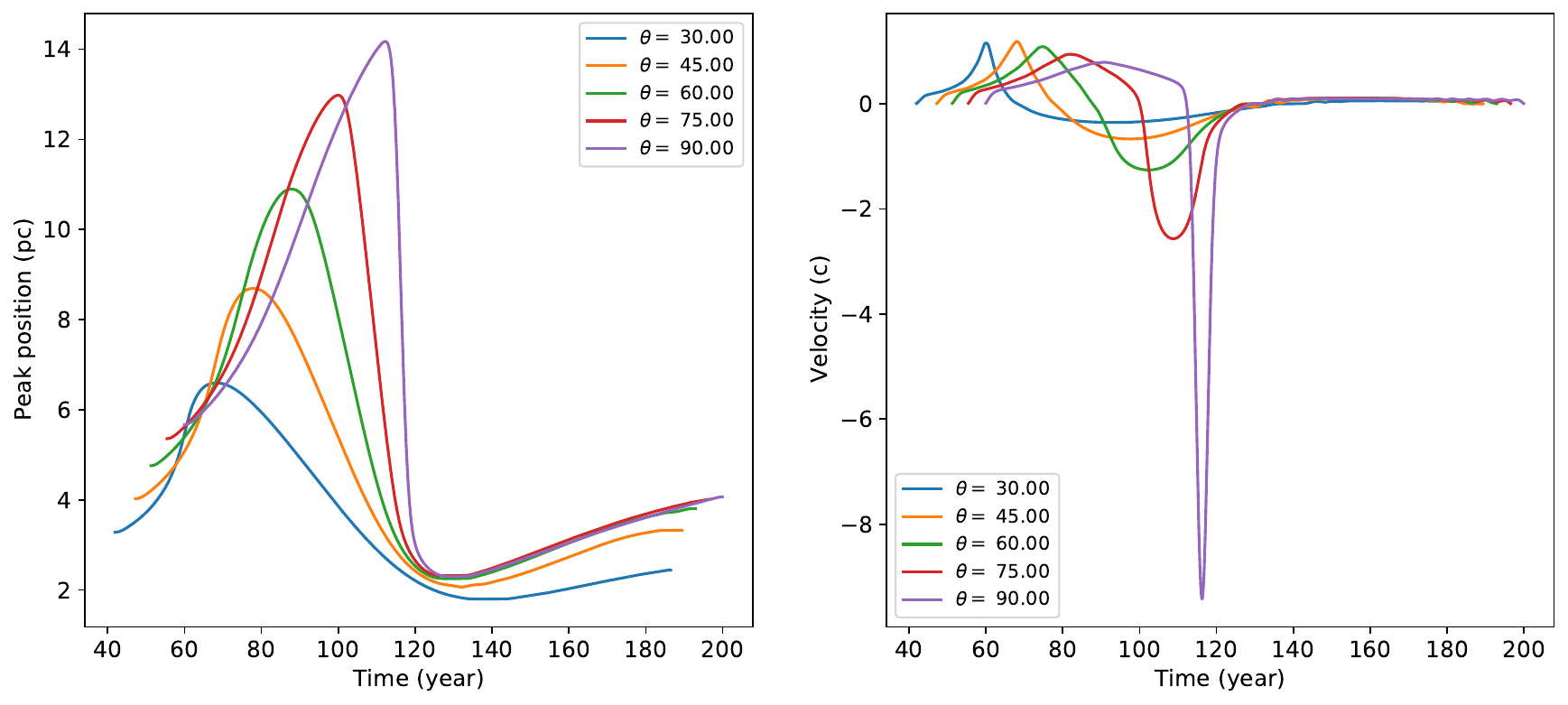}
    \caption{Left: evolution of position of peak emissivity with time, for different viewing angles $\theta$ for $\beta_{\rm post}=0.6$ and $\alpha=0.5$. As expected, the time intervals when the knot is moving toward us are reduced and that when the knot is moving away (inward) are increased. Right: the corresponding apparent velocity as a function of time. The maximum backward speed is reached for $\theta=90^\circ$, when it decreases for decreasing $\theta$, since the time intervals between successive pulses from the same knot is dilated since it is moving away from us. The maximum inward speed for $\theta=90^\circ$ represents $\beta_{\rm post}$.}
    \label{fig:theta}
\end{figure*}

\begin{figure*}
    \centering
        \includegraphics[width=\linewidth]{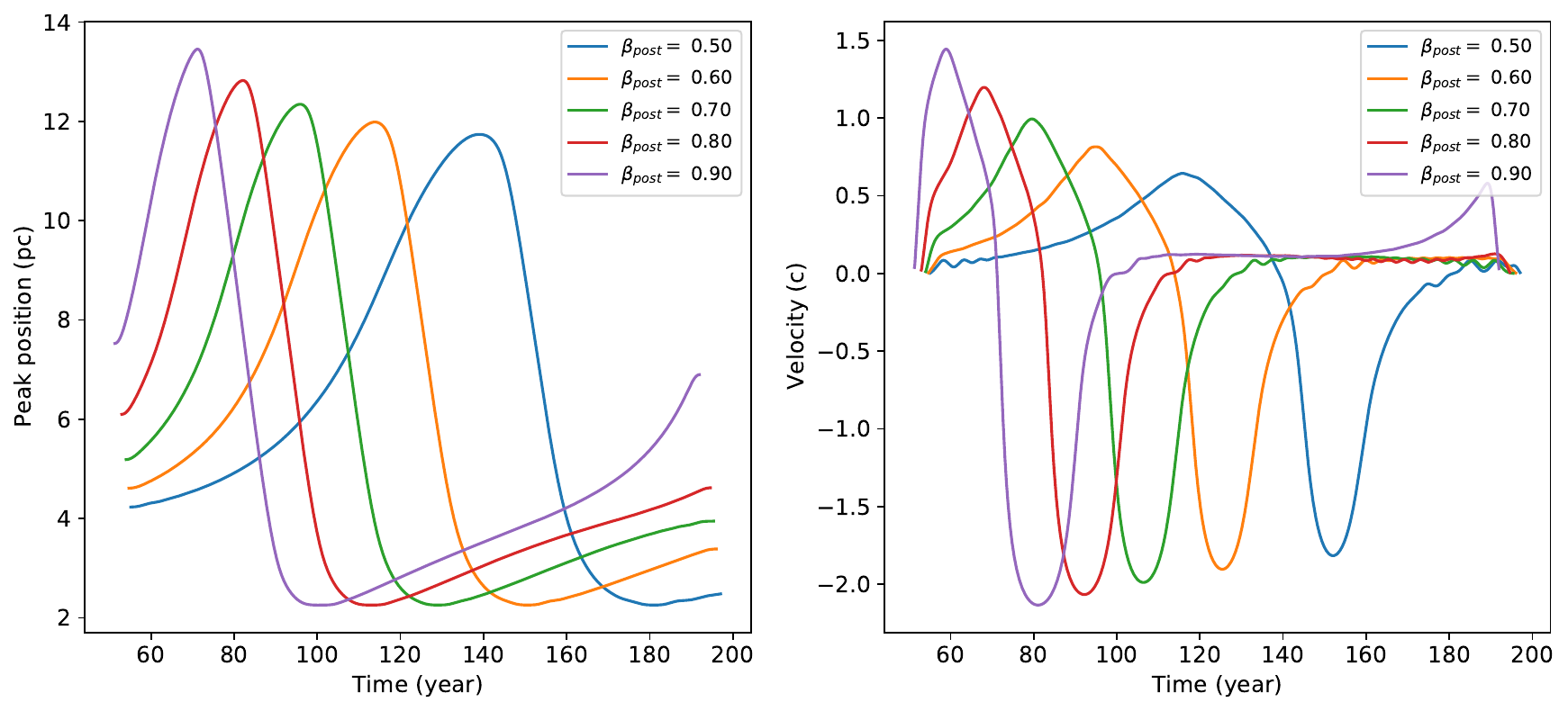}
    \caption{Left: evolution of position of peak emissivity with time, for different $\beta_{\rm post}$ at $\theta=70^\circ$ and $\alpha=0.5$. Lower $\beta_{\rm post}$ slows the duration of increase of the peak position, but decaying half is observed to have a slope, nearly independent of $\beta_{\rm post}$. Right: the corresponding apparent velocity as a function of time. The maximum backward speed is almost insensitive to $\beta_{\rm post}$, but the outward speed sharply increases due to increasing $\beta_{\rm post}$, because that is mainly driven by $\beta_{\rm post}$.}
    \label{fig:bpost}
\end{figure*}

\begin{figure*}
    \centering
    \includegraphics[width=\linewidth]{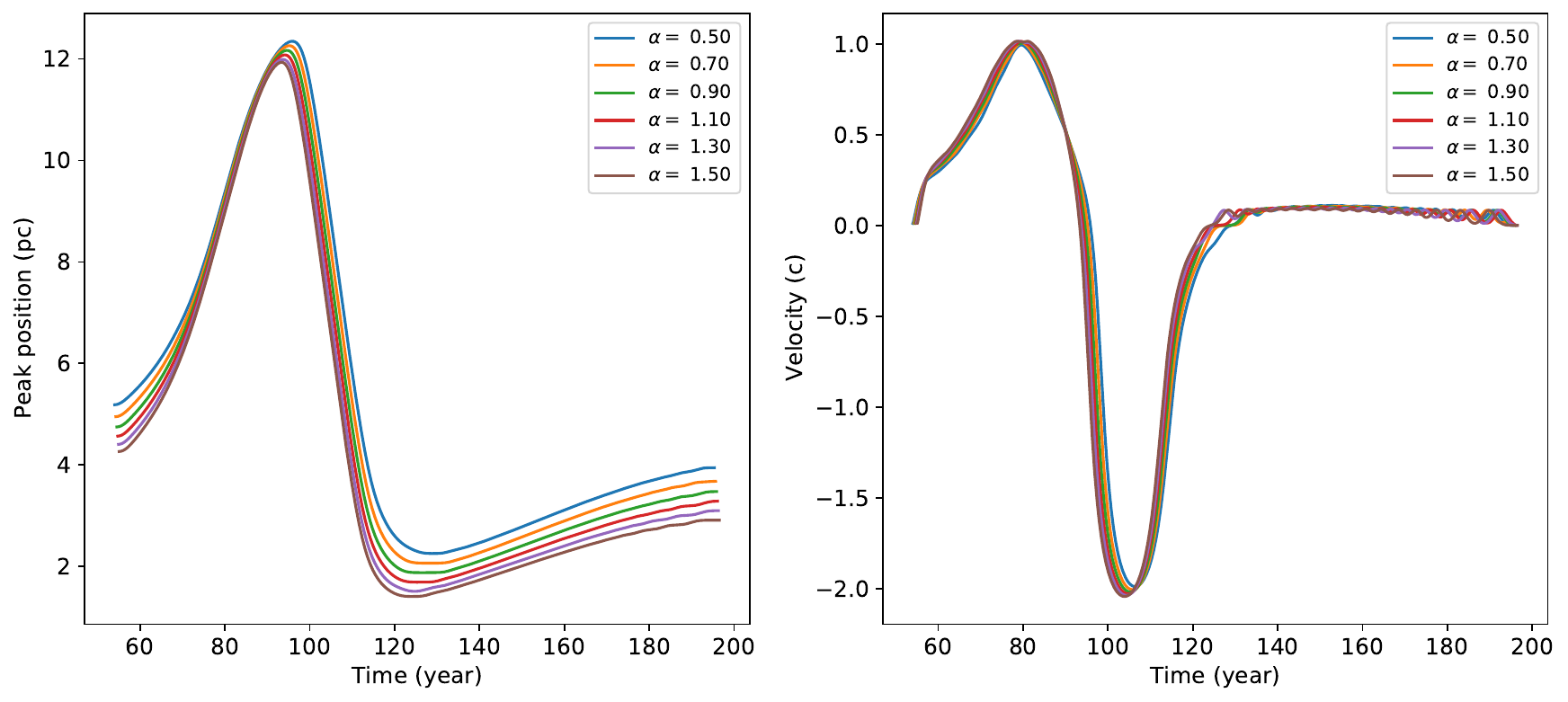}
    \caption{Left: evolution of position of peak emissivity with time, for different $\alpha$ for $\beta_{\rm post}=0.6$ at $\theta=70^\circ$. Right: the corresponding apparent velocity profile. Both the plots shows clear negligible dependence of $\alpha$ on the peak position evolution and its velocity, over a physically plausible range.}
    \label{fig:alpha}
\end{figure*}

For a conical profile given as $r\propto z$ (see Section \ref{sec:profile}), $\alpha=0.5$, $z_f=z_s+\beta_{\rm post}ct$ with $\beta_{\rm post}=0.6$ and $z_s=40$ pc, we plot the corresponding peak position $z_{\rm max}(z,t)$ and the resulting velocity $\beta_{\rm app}(z,t)$ for different values of $\theta$ at different times $t$, as a function of $z$ in Figure \ref{fig:theta}. While it is clear that with decreasing $\theta$, the velocity of outward motion increases (due to time contraction) and that for inward motion decreases (due to time dilation), the inward motion for fairly large viewing angles $\theta\gtrsim\pi/3$ is superluminal. Furthermore, the traversed spatial scales are few to tens of parsecs over decades, akin to our case here. The same cycle of forward and backward motion continues as $\alpha_p$ keeps increasing with $z$. For $\theta=90^\circ$, the maximum outward speed is essentially $\beta_{\rm post}$, reached right before a superluminal backward motion. Figure \ref{fig:bpost} shows the same ($z_{\rm max}(t)$ and $\dot{z}_{\rm max}(t)$) for $\theta=70^\circ$, $z_s=40$ pc. Slower $\beta_{\rm post}$ evidently results in a much slower forward and backward motion. For example, for $\beta_{\rm post}=0.9$, the total duration of rise and fall of the peak position is less than half that of $\beta_{\rm post}=0.5$. On the other hand, while the maximum observed backward speed is lightly sensitive to $\beta_{\rm post}$, the maximum outward speed is more sensitive to $\beta_{\rm post}$ as expected, and increases strongly with the latter. We show a similar illustration, but for different values of $\alpha$ given $\beta_{\rm post}=0.6$ and $\theta=70^\circ$, in Figure \ref{fig:alpha}. As clear, the evolution of the peak position, and thereafter its velocity, is negligibly sensitive to $\alpha$. While higher values of $\alpha$ imply a faster adiabatic expansion, as in Equation \ref{eq:jnu}, Figure \ref{fig:alpha} implies that a combined effect of $(\sin\alpha_p)^{\alpha+1}$ and $z^{-4\alpha}$ weakens the dependence of the peak position and velocity on $\alpha$.

\end{document}